## Title

Viscous Dissipation Governs Bubble Morphology and Failure in Soft Matter

## Authors

Sushma SP[1#], Anshul Shrivastava[1#], Imnatoshi Jamir[2], Aritra Chatterjee[3], Namrata Gundiah[1*]

## Affiliations

[1]Department of Mechanical Engineering, [2]Centre for Nanoscience and Engineering, Indian Institute of Science, Bangalore, India.

[3]Jyoti & Bhupat Mehta School of Health Sciences & Technology, Indian Institute of Technology Guwahati, Assam, India.

# Equal Contribution

* For correspondence:

Namrata Gundiah, Biomechanics Laboratory, Mechanical Engineering Department, Indian Institute of Science, Bangalore 560012.

Email: namrata@iisc.ac.in; ngundiah@gmail.com

Tel: 91 80 2293 2860/ 3366

## Abstract

Cavitation, the growth of bubbles in fluids and soft matter, plays a central role in tissue damage, ultrasound therapies, and material failure, yet the influence of viscoelastic dissipation on bubble dynamics remains unclear. Here, we investigate cavitation in polyacrylamide hydrogels with identical elastic moduli but different viscous dissipation. We observe a striking symmetry-breaking transition: elastic gels rapidly develop ellipsoidal cavities before rupture, whereas viscoelastic gels sustain large, nearly spherical bubbles. A modified Rayleigh–Plesset framework shows that viscous stresses suppress shape instabilities and delay symmetry breaking, whereas elastic stresses accelerate cavity deformation. Finite element simulations independently validate these findings. The Deborah number (De) further captures the competition between material relaxation and cavity growth, with higher De associated with predominantly elastic growth and lower De with greater relaxation. Together, our results establish viscous dissipation as a key stabilizing mechanism that governs bubble morphology and failure in soft materials. These results have direct implications for understanding cavitation-mediated damage in biological tissues, optimizing therapeutic ultrasound, and advancing mechanical characterization of hydrogels for engineering and food processing applications.

## Significance

Cavitation is viewed as an instability, where bubbles grow uncontrollably until failure, causing damage in biological tissues during blast injuries, ultrasound medical therapies, and turbine blades. Elasticity resists bubble growth, whereas the role of viscous dissipation is less well studied. We show that viscous dissipation stabilizes and delays the onset of shape instabilities, altering the rupture pathway. By engineering hydrogels with matched elasticity but varied viscosity, we show that bubble dynamics are governed by a competition between the surrounding elasticity and viscosity rather than a purely elastic phenomenon. The physical framework in our study predicts these transitions and offers a method in biomaterials design and in optimizing clinical ultrasound to minimize collateral tissue damage. These findings redefine our understanding of soft matter under extreme mechanical stress.

# MAIN TEXT

## Introduction

Micron-sized bubbles arise in diverse settings, from decompression in blood vessels to cavitation in naval and turbomachinery applications, where their rapid growth and collapse can produce severe structural damage (1,2). Beyond these contexts, bubble cavitation plays an important role in industrial processes such as food processing (3). In biomedicine, rapidly oscillating microbubbles are widely used as ultrasound contrast agents (4) and as therapeutic tools for targeted drug delivery and gene therapy *via* ultrasound-induced collapse (5). Across all such systems, bubble dynamics are governed by the mechanical properties of the surrounding medium. Although traditionally associated with fluids, cavitation is now understood to play an important role in damage initiation and failures of soft materials, including biological tissues (1). As soft materials find widespread use in engineering, biomedical, and tissue engineering applications, an understanding of how damage initiates and propagates within soft materials has become an important area of research. Addressing these questions requires approaches that integrate insights from diverse fields, including mechanics and physics, biology and chemistry.

Three-dimensional, water-swollen polymeric hydrogel networks provide a model platform for investigating cavitation in soft matter because they can be engineered to exhibit low moduli while withstanding large compressive deformations, thus mimicking key features of biological tissues. The viscoelastic properties of such materials emerge from physical crosslinks between polymer chains, chain entanglements, and fluid-network interactions, which together generate dissipative and time-dependent mechanical responses. Despite spanning a broad range of elastic moduli (1 Pa to 10 MPa) (6), synthetic single-network hydrogels typically exhibit low fracture toughness (~10 J/m$^2$) as compared to natural water-swollen tissues such as cartilage and cornea that exhibit toughness values approaching ~1000 J/m$^2$ (7-13).

Pressure-induced cavitation rheology is a powerful technique for probing the local mechanical properties of hydrogels and tissues by measuring the internal pressure within a bubble during expansion (14-17). In this method, air is injected through a blunt-tipped needle to introduce a flaw of known size, and the critical pressure ($P_c$) for bubble nucleation is recorded (1). The critical energy release rate, $G_c$, of the material is estimated using $P_c$ and by assuming neo-Hookean behavior under quasi-static inflation (16-18). A variation of this technique, laser induced cavitation, uses a spatially-focused pulsed laser to generate inertial bubbles at high strains (1, 19). Microbubbles act as highly sensitive mechanical probes of the local milieu of soft materials, and exhibit complex and often nonlinear behaviors. Acoustic measurements of microbubble dynamics also provide a powerful approach to characterize hydrogel viscoelasticity and failures (20, 21). Yet, bubble collapse dynamics remain exquisitely sensitive to the viscoelastic properties of the surrounding medium (22), and the fundamental mechanisms governing bubble growth and collapse in soft, highly deformable hydrogels remain poorly understood. Controlling and manipulating bubble dynamics in viscoelastic materials remains challenging, yet is critical for applications such as targeted drug delivery, where precise localization of the therapeutic is essential, and for minimizing collateral damage to surrounding healthy tissues during ultrasound.

Hydrogels with matched elastic moduli, though differing in their energy dissipation characteristics, provide valuable model systems to isolate the role of viscous dissipation in bubble growth and collapse. We hypothesize that enhanced dissipative effects in viscoelastic hydrogels promote the formation of larger, more spherical bubbles, whereas elastic hydrogels, with lower viscous dissipation, exhibit smaller and less spherical bubble shapes. To test this hypothesis, we fabricated elastic (E) and viscoelastic (VE) polyacrylamide (PAAm) hydrogels with comparable elastic moduli but significantly different viscous losses (23), and investigated needle-induced cavitation across a range of inflation rates. Bubbles grew larger and retained a spherical morphology in the viscoelastic PAAm hydrogels, whereas those in the elastic gels initially formed as spheres but transitioned to elliptical shapes prior to collapse. We analyzed these symmetry breaking transitions using a modified Rayleigh-Plesset equation (24) that incorporates both elastic and viscous stress contributions during bubble growth. We demonstrate the importance of relaxation times that contribute to the differences in bubble shapes in the elastic and viscoelastic hydrogels. We also calculated the Deborah number (De) which captures the competition between material relaxation and cavity growth. Greater energy dissipation in VE hydrogels produced larger, more spherical bubbles with roughened surfaces, whereas dominant elastic stresses in E hydrogels promoted shape asymmetry and instability. These results were consistent with finite element simulations of bubble expansion in the E and VE hydrogels that allowed us to quantify the distinct contributions of elastic and viscous stresses during bubble expansion. These findings demonstrate the role of rheological properties in governing bubble-matrix interactions, dictating whether collapse results in uniform or axis-localized damage. Asymmetric bubble growth mechanisms enable a better understanding into how cavitation-induced stresses may contribute to trauma and impact-related injuries, providing a framework to bridge fundamental aspects of bubble dynamics with biomedical and protective applications. Such insights are specifically relevant to enhancing the safety and efficacy of biomedical interventions, including ultrasound-mediated drug delivery.

## Results and Discussion

### Mechanical properties of elastic and viscoelastic PAAm hydrogels

We fabricated PAAm hydrogels with similar elastic properties, but significantly distinct viscoelastic contributions by incorporating long linear acrylamide polymer during the gel formulation (Fig. 1A, B). Details of the fabrication protocol are described elsewhere; a brief description is provided in Methods (23). In this study, we refer to hydrogels with enhanced viscous components as “viscoelastic” hydrogels (VE) and those with lower viscous contribution as “elastic” (E). The elastic moduli were remarkably similar between the VE (13.09 ± 1.16 kPa) and E (12.58 ± 0.62 kPa; p>0.33) hydrogels in our earlier study (23). Dynamic Mechanical Analysis (DMA), also reported in our prior study (23), showed that the loss modulus of VE gels was 36.9% higher than E gels at 0.05 Hz; this difference increased to 50% at 70 Hz. Monotonic compression experiments showed that both E and VE hydrogel samples exhibited large nonlinear stress-strain behaviors (Fig. 1C, D). We fit these data to a neo-Hookean strain energy function (Equation 4), and modeled the responses using finite element (FE) simulations in ABAQUS (Dassault Systèmes Simulia Corp., Providence, RI, 2022) using the optimized $C_{10}$ parameter (see Methods; Fig. 1E; Table 1; Fig.S1). FE simulations using the neo-Hookean model showed good agreement with experimental results for both E and VE samples (Fig. S2).

To test the viscoelastic properties, we performed stress relaxation experiments and modeled the time-dependent stress decay under a fixed strain input using a phenomenological Maxwell-Weichert model. This model includes a linear spring in parallel with two Maxwell elements (Fig. 1F), and captures the relaxation processes occurring over distinct time scales that are represented by characteristic times, $\tau_1 = \frac{{}_1}{{}_1}$ and $\tau_2 = \frac{{}_2}{{}_2}$, where $E_1, E_2$ and $\eta_1, \eta_2$ denote the spring elastic moduli and dashpot viscosities of the two Maxwell elements, respectively. The fast relaxation time, $\tau_1$, was similar for both hydrogels (9.8 min for E hydrogels and 9.4 min for VE hydrogels). In contrast, the slow relaxation time constant, $\tau_2$, differed substantially between the groups: $\tau_2$=133.1 min for E gels whereas the corresponding value was 90.4 minutes for VE hydrogels. The lower relaxation value observed in the VE group indicates greater energy dissipation, that may arise from the presence of linear PAAm within the hydrogel network.

We incorporated these time-dependent behaviors into the FE simulations using a Prony series-based viscoelastic material model (24). The phenomenological Maxwell-Weichert model is mathematically equivalent to a two-term Prony series representation of linear viscoelasticity. The total relaxation modulus is given by

$$E(t) = E_\infty + E_1 e^{\frac{-t}{\tau_1}} + E_2 e^{\frac{-t}{\tau_2}} \tag{1}$$

$E_\infty$ is the long-term equilibrium modulus, and $\tau_1, \tau_2$ are relaxation times for the two Maxwell elements in the model (Methods). The moduli and relaxation times from the Weichert fit were used to compute the Prony parameters $(g_i, \tau_i)$ in the FE model. The Prony series coefficients (Table 2) showed good agreement with those obtained from fitting the experimental stress relaxation data (Fig. 1G).

**Bubble growth in elastic and viscoelastic hydrogels**

We used needle-induced cavitation rheology to investigate bubble dynamics in the E and VE hydrogels at controlled inflation rates of 500, 1000, 2000, 4000, 6000 and 8000 µl/min (Fig. 2A). Pressure in the cavity was monitored using a transducer, and bubble shape evolution was recorded using an overhead camera (17). Representative pressure traces for E and VE hydrogels at 2000 µl/ min (Fig. 2B) show a characteristic increase in pressure, followed by unstable bubble growth and subsequent collapse. This irreversible bubble growth and collapse results in hydrogel fracture and is defined as failure in this study. The pressure increased linearly with time until the point of bubble collapse across all inflation rates in the study at 500, 1000, 2000, 4000, 6000 and 8000 µl/ min for specimens in the E and VE groups, respectively (Fig. 2C). We used these data to obtain the onset of instability and the corresponding critical pressure, $P_c$. Although there were no significant differences in $P_c$ across all inflation rates in the two hydrogel groups in our study, the VE gels consistently exhibited significantly higher $P_c$ values compared to corresponding $P_c$ values for the E hydrogels (Table 3, $p<0.01$) at each inflation rate. Additionally, the time for bubble initiation varied inversely with inflation rate, indicating a rate-dependent onset of cavitation (Fig. 2D). The overall similarity in rate dependence for E and VE hydrogel groups demonstrates predominantly elastic hydrogel behaviors under cavitation rates tested in our study. We next used the measured $P_c$ values, to compute the fracture toughness ($G_c$) of the hydrogels using a linear elastic formulation. The VE hydrogels showed significantly higher $G_c$ compared to the corresponding values computed for the E group for all inflation rates tested in the study (Fig. 2E; Table 3; $p < 0.01$), suggesting enhanced hydrogel resistance due to the presence of linear acrylamide. These results highlight the role of viscous dissipation in hydrogel failure, and demonstrate how rheological properties modulate cavitation failures.

Previous studies have demonstrated that the expression for the critical energy release rate, ($G_c$), associated with cavitation in a neo-Hookean material becomes analytically unbound (25-27). Thus, an LEFM-based framework has been used to estimate the fracture toughness of hydrogels from cavitation experiments. Although cavitation involves large deformations and is inherently nonlinear, this approach has been shown to provide a reliable first-order estimate of fracture toughness for neo-Hookean and strain-hardening polymeric materials (28, 29). Consistent with these findings, our previous work in gelatin hydrogels demonstrated a good agreement between fracture toughness, estimated from cavitation using the LEFM framework, and that obtained from nonlinear fracture tests, including single-edge-notch and pure-shear-notch tests (30).

Fig. 3A and B show representative images of bubbles in the E and VE hydrogels prior to collapse at each of the different inflation rates in our study. Bubbles in VE hydrogels displayed longer lifetimes before collapse compared to those in E gels (Table 3). We also observed pronounced differences in their morphology prior to collapse: bubbles in VE gels were larger and more spherical, whereas those in the E gels were smaller and ellipsoidal across all inflation rates tested in our study (Supplementary Movie 1-12). To quantify these morphological differences, we measured the effective bubble radius, $R_{eff}$, and aspect ratio (AR) using image analysis (see Methods). $R_{eff}$ was consistently higher in the VE gels compared to E gels at each inflation rate (Fig. 3C), and increased non-linearly with time following initiation until final bubble collapse. Slope of the $R_{eff}$ – time curve was also steeper in the VE gels, indicating faster bubble growth relative to the E hydrogels. AR changes showed distinct variations in the two hydrogel groups: AR started significantly greater than 1 during bubble growth in the VE gels (Fig. 3D) and decreased to values nearing unity, indicating a transition from an initial ellipsoidal to nearly spherical shape during collapse. Bubble growth in the VE hydrogels was spherical until failure at all inflation rates tested in the study (Fig. 3). In contrast, bubbles in the E hydrogels exhibited higher AR values at initiation, and decreased to a minimum before gradually increasing again prior to collapse. These results indicate a reversion of the bubble geometry towards an ellipsoidal shape. The inflection point in AR indicates the time transition for the symmetry change in E hydrogels, where the bubble transitions from spherical to ellipsoidal geometry (Table 3). These data indicate that whereas all bubbles begin as ellipsoidal shapes, their subsequent evolution is governed by the rheological properties of the surrounding medium.

The transition from spherical to penny-shaped cavity growth in soft materials is characterized by the elasto-adhesive length, which scales with the ratio of fracture energy ($G_c$) to elastic modulus (25-27). In our study, the initial defect size (150 μm) exceeds the elasto-adhesive length of the E gels, suggesting the formation of penny-shaped cavities. This value is however smaller than that of the VE gels (Fig. 3E) that show spherical cavity growth. These predictions agree with our experimental observations.

The transition from cavitation to fracture is similarly characterized by the elasto-cohesive length, given by $\frac{G_c}{W_f}$, where $W_f$ is the work of fracture obtained from the area under the uniaxial stress–stretch curve up to failure (31). The estimated elasto-cohesive lengths for all gels in our study were smaller than the initial defect size (Fig. 3F), suggesting that fracture accompanies cavity growth which is consistent with the rough cavity surfaces observed experimentally. Because $W_f$ was not measured in our experiments, we estimated it from polyacrylamide gels of comparable modulus reported by Fu and colleagues (31). These characteristic lengths should hence be regarded as approximate. This limitation is particularly relevant for VE gels, where strain-dependent material properties can limit the quantitative interpretation of such length scales (29).

Early studies by Gent and coworkers established the occurrence of cavitation in vulcanized rubber cylinders during tension tests: the bubble expands and becomes unstable under sufficiently negative hydrostatic stress (14). Building on this foundation, Zimberlin and coworkers developed cavitation rheology as a quantitative technique to relate the critical cavitation pressure to the fracture toughness of soft materials using a linear elastic formulation (15, 16). Subsequent studies inferred from cavitation measurements showed that hydrogel properties, including elastic modulus and fracture toughness, depend on the rate of bubble inflation (32, 33). Theoretical and experimental studies further emphasize the importance of viscosity and viscoelasticity in cavitation behaviors. Qin and coworkers developed a model demonstrating that material viscosity dampens bubble dynamics and reduces the corresponding collapse strength (34). Viscosity of the surrounding non-Newtonian medium induces oscillations during bubble collapse which has a mitigating effect on cavitation damage (35). Models also demonstrate that nonspherical instabilities during cavitation in soft nonlinear solids strongly depend on the material viscoelasticity (36). Despite these insights, few studies have delineated the individual contributions of elastic and viscous components in cavitation rheology. Addressing this gap is critical, as many biological and engineered hydrogels are inherently viscoelastic, and their resistance to cavitation may depend on both the inherent dissipation and elastic responses. Experiments from our study show that viscoelasticity delays symmetry breaking and promotes spherical cavity expansion, whereas elasticity favors an early transition to asymmetry. The marked differences in bubble morphology and dynamics suggests the crucial role of viscous dissipation in shaping cavitation behaviour in soft hydrogels.

**Role of elastic and viscous stresses during bubble expansion in hydrogels**

Cavitation dynamics of a single gas-filled bubble in a Newtonian fluid are classically described using the Rayleigh-Plesset equation - a second order nonlinear differential equation describing the evolution of bubble radius with time (1, 37, 38). Bubble growth is governed by interface motion induced by pressure differences across the cavity, assuming negligible interfacial mass transport and the absence of body forces on the timescales of bubble growth and collapse. To account for the mechanical properties of biological tissues and hydrogels, Gaudron and colleagues (39) modified the Rayleigh-Plesset equation to include the effects of the surrounding elastic and viscous stresses (Equation 14). We adopt this modified formulation to model the dynamics of bubble growth in E and VE hydrogels in our study. This approach enables direct comparison with the experimental observations of symmetry-breaking transitions in our work.

Because of the challenges in obtaining closed-form solutions to the modified Rayleigh-Plesset equation, we analyzed the experimental form of bubble growth evolution in the E and VE hydrogels during inflation and used these to develop empirical evolution equations for the growing bubble radius and the corresponding radial velocity (Equations 12, 13). Figures 4A and B show representative fits of $R_{eff}$ with time in the E and VE hydrogels at inflation rates of 500 and 8000 µl/min, respectively, using the empirical growth law (Equation 12). The initial radius of the bubble ($R_0$), corresponding to the bubble size at the beginning of the inflation process and the associated values of $R_1$ and t were consistently lower for E gels compared to the VE group (Table 4), demonstrating smaller bubbles in E gels as compared to the VE gels. Changes in the radial velocity, $\frac{dR_{eff}}{dt}$, calculated using Equation 13, are shown in Figures 4C and D, respectively. The observed decay in velocity suggests that bubble expansion is initially driven by flow, but slows down over time due to increasing resistance due to gel elasticity and viscous dissipation. Similar results were obtained for variations in $R_{eff}$ and $\frac{dR_{eff}}{dt}$ for the other inflation rates tested in the study (Fig. S3).

To investigate the relative importance of elastic and viscous stresses in bubble growth, we developed a 2D FE model using constants from the neo-Hookean model based on compression test simulations and the two-term Prony series viscoelastic material simulations (see Methods; Fig. 5A). Mesh refinement was applied in the vicinity of the cavity, where element sizes were significantly smaller than those in remote regions significantly away from the cavity (Fig. 5A). The radial displacement history due to cavity expansion (Equation 12) obtained from experimental measurements were used as inputs to the model for all inflation rates in the E and VE hydrogel groups. The elastic stress term was determined from FE simulations of neo-Hookean elastic material. Viscous stresses were obtained from the FE simulations using the bubble radius and radial velocities using the expression $\frac{4\nu_L \dot{R}}{R}$. We also obtained the individual stress contributions from elastic, viscous and surface tension terms using the modified Rayleigh-Plesset equation for each of the inflation rates using the radial stress variations. FE results showed strong agreement with the analytical model as shown in results from 500 µl/min inflation for a representative sample in the E hydrogel (Fig. 5B) and VE hydrogel groups (Fig. 5C), respectively. Notably, the E gels exhibited rapid reduction in viscous stresses, whereas the VE gels showed gradual decay for these inflation rates. Comparable results for the elastic and viscous stresses were obtained at the other inflation rates tested in our study. These results agree with data supporting longer bubble growth durations in VE gels compared to the smaller and faster bubble collapse in the E hydrogels.

We used the experimentally obtained bubble radius and growth velocity evolution to compute a kinetic energy related growth velocity gradient (GVG) term, $R\ddot{R} + \frac{3}{2}\dot{R}^2$, in the modified Rayleigh-Plesset equation. The GVG exhibited a sharp minimum (~33.34 ms) in E hydrogels at approximately the same time as the inflection point in the AR variation (Fig.S5; Table 3). In contrast, VE hydrogels showed a more gradual and monotonic decay in GVG and AR values, consistent with a more sustained spherical morphology during bubble growth (Fig.S6). The drop was sharper at higher inflation rates as compared to that seen at 500 µl / min. This transition aligns with the assumptions underlying the Rayleigh-Plesset model that describe spherical bubble growth in a viscous medium. Together, these findings show that bubble morphology and consequent evolution are tightly coupled to the mechanical response of the surrounding hydrogel material.

To assess the interplay between material relaxation and cavity growth timescales, we calculated the Deborah number ($De = \frac{\tau_2}{\frac{R}{\dot{R}}}$) and compared it with the cavity aspect ratio (AR) for all inflation rates (Fig. 6). For E gels, AR was evaluated at the inflection point, whereas for VE gels, which lacked a distinct inflection point, the AR was evaluated when the graph reached a constant value. The E gels generally exhibited higher De values than VE gels, indicating that cavity growth occurs on a shorter timescale relative to stress relaxation. In contrast, the lower De values of VE gels show that stress relaxation occurs simultaneously with cavity growth, resulting in a greater influence of viscoelastic relaxation on the cavity morphology.

To delineate the role of elasticity in determining the minimum bubble radius during cavitation, we use the Wright-omega functions and compared the predicted values with the initial bubble radius, $R_o$, in the modified Rayleigh-Plesset model (39). Specifically, for a material with density, $\rho$, the ratio of minimum to initial bubble radius is given by

$$\frac{R_{min}}{R_0} = \frac{e^{[W(log\left(-\frac{B}{A}\right)+(C/A)]}}{e^{(C/A)}} \tag{2}$$

where, A=$p_{Go}/\rho$ , B=$-\mu/\rho$ and C=(2[Δp]/3 $\rho$) +(2S/ $\rho R_0$) +( $2\mu/3\rho$). In the absence of the elastic shear modulus ($\mu$), this expression simplifies to

$$\frac{R_{min}}{R_0} = e^{\frac{A}{C}} \tag{3}$$

The material parameters and the bubble radius growth law were incorporated in this framework to calculate $R_{min}$ and quantify the contribution of elasticity to the minimum bubble radius obtained prior to collapse. These values showed good agreement with the fitted $R_o$ values from our experimental data (Table 4). Comparisons between experiments performed at other inflation rates for the E and VE gels with the modified Rayleigh-Plesset model simulations are shown in Fig. S7. The minimum bubble radius at collapse was consistently larger for VE hydrogels compared to E gels and agreed with trends observed in the experimental data. Together, these results support the applicability of the modified Rayleigh-Plesset framework to study bubble dynamics in E and VE gels. They also highlight how material elasticity resists collapse by reducing not only the bubble acceleration but also providing the recoil force. In contrast, the dominant effect of viscosity in VE hydrogels inhibits shape deformation, to maintain the spherical bubble geometry across all inflation rates in our study.

## Conclusions

We investigated the role of differential viscoelasticity in cavitation dynamics using PAAm hydrogels engineered to exhibit similar elastic moduli but with distinct viscous contributions. Elasticity in the hydrogels arises from crosslinked polyacrylamide networks, whereas viscous responses are attributed to water mobility and the presence of un-crosslinked linear PAAm in the network structure. By integrating experiments with numerical simulations, we identify how viscous dissipation alters bubble growth morphology and shapes, thereby providing a framework to understand symmetry breaking transitions in bubble shapes in hydrogels. We used stress relaxation experiments to characterize the viscoelastic response associated with the presence of linear acrylamide in the engineered hydrogels. Our results demonstrate that VE samples exhibited a lower relaxation time ($\tau_2$) than E hydrogels, resulting in higher energy dissipation. Cavitation rheology studies performed over a range of inflation rates (500-8000 µl/min) showed higher critical pressures ($P_c$), and correspondingly greater fracture toughness (Gc) of VE hydrogels compared to their elastic counterparts. Distinct differences were observed in bubble morphology: bubbles were highly ellipsoidal in E hydrogels as compared to spherical shapes in VE hydrogels. Although all bubbles were initially spherical, those in VE hydrogels retained spherical symmetry whereas bubbles in E hydrogels transitioned into increasingly elliptical shapes. Using a modified Rayleigh-Plesset framework (37-39), coupled with finite element analyses for cavity expansion in viscoelastic media, we show that elastic stresses dominate over viscous dissipation and drive the shape asymmetry, whereas viscous stresses suppress shape perturbations and promote symmetry during cavity expansion in hydrogels. Viscosity however, inhibited the shape deformations and helped maintain bubble symmetry during cavity expansion. Higher De values in E gels indicate predominantly elasticity driven cavity growth, whereas the lower De values in VE gels highlight the greater influence of stress relaxation on cavity morphology. Our results further indicate that material elasticity resists bubble collapse by limiting bubble acceleration and providing a restoring recoil force. On the other hand, viscosity suppresses shape distortions but introduces oscillations in the evolving bubble shapes. Surface tension effects were negligible compared to the elastic and viscous contributions to the stresses in cavitation rheology (Fig. S8). Asymmetric bubble expansion and collapse in elastic hydrogels produce non-uniform, axis-localized damage, which has implications for cavitation induced injury in biological tissues and other soft materials. In conclusion, transitions from spherical to elliptical bubble shapes may significantly alter the spatial distribution of stresses in cavitation-based processes in living tissues and soft matter. These results pave the way towards a better understanding into how cavitation dynamics govern damage localization in viscoelastic materials.

## Materials and Methods

### Uniaxial mechanical experiments and stress relaxation of hydrogels

Polyacrylamide (PAAm) hydrogels were fabricated by crosslinking acrylamide with bisacrylamide using ammonium persulfate (APS) and tetramethylethylenediamine (TEMED) based on previously reported protocols (23). Elastic hydrogels were prepared by mixing acrylamide stock solution (40% w/v) with bisacrylamide stock solution (2% w/v) in deionized (DI) water. For VE hydrogels, a highly viscous linear polyacrylamide (linear PAAm) solution was first synthesized and cured at 37 °C for 1 hour. Linear PAAm was blended with acrylamide and bisacrylamide stock solutions used for the elastic hydrogels (Fig. 1A, B). Both formulations of elastic and viscoelastic mixtures were cast into cylindrical molds (16 mm in diameter, 10 mm in height), and polymerization was initiated by adding APS (10% w/v) and TEMED (Sigma Aldrich, T9281). The mixtures were gently stirred, degassed, and allowed to polymerize at room temperature for 30 minutes. Cured specimens were removed from the molds and stored in DI water at 4°C until further use.

Monotonic compression tests were performed on the E (n=4) and VE (n=4) hydrogel samples as described earlier (17). Briefly, the specimen was placed between parallel compression platens in a Bose ElectroForce® 3200 Dynamic Mechanical Analysis (DMA) system and preloaded to 10 gm. To ensure a repeatable material response, samples were preconditioned by cycling for 10 cycles of 5 % strain at 0.05 Hz, followed by quasi-static compression at 0.01 mm/s until failure (40). Forces were recorded using a 25 N transducer (Honeywell Sensotec Inc., Columbus, OH, USA) and used to compute engineering stress based on the initial cross-sectional area. Engineering strain was calculated from the change in the sample height during compression.

We used a neo-Hookean strain energy function, $\psi$, to represent the constitutive properties of the specimens, given by:

$$\psi = C_{10}(I_1 - 3) \quad (4)$$

$C_{10}$ $(kPa)$ is a material constant. $I_1$ is the first invariant of the left Cauchy-Green deformation tensor (**B**), $I_1(B) = tr(B)$. Assuming material incompressibility, we write $I_3 = (\lambda_1\lambda_2\lambda_3)^2 = 1$, where $\lambda_i; i = 1:3$ are the stretches along the principal directions. The corresponding Cauchy stresses are obtained as

$$\sigma_i = \quad {}_i\left(\frac{\partial\psi}{\partial\lambda_i}\right) - p \quad (5)$$

where p is a Lagrange multiplier and corresponds to a pressure term. For uniaxial compression, the stretches along the principal directions are given by $\lambda_1 = \lambda$. Using incompressibility, we write $\lambda_2 = \lambda_3 = 1/\sqrt{\lambda}$. Assuming traction-free sides, $\sigma_{22} = \sigma_{33} = 0$, the Cauchy stress in the loading direction for a neo-Hookean material, is given by

$$\sigma_{11} = 2C_{10}\left(\lambda^2 - \frac{1}{\lambda}\right) \quad (6)$$

Experimental stress-strain data were fit to a neo-Hookean strain energy function in MATLAB (R2020b, Natick, MA) using the function *lsqnonlin*. Goodness of fits were determined using Root Mean Squared error (RMSE) values given by

$$\text{Root mean squared error (RMSE)} = \sqrt{\frac{\Sigma_i^N \quad (\sigma_{11(NH)} - \sigma_{11\,(Exp)})_i^2}{N}} \quad (7)$$

Quasi-static stress relaxation experiments were conducted to quantify the viscoelastic behaviors of the elastic and viscoelastic PAAm hydrogels under unconfined compression. Cylindrical samples were preloaded to 10 grams and preconditioned at 5 % strain for 10 cycles at 0.05 Hz. Specimens were strained to 10% and held for 4 hours. Temporal variations in loads were recorded during this duration to obtain stress relaxation data for the E and VE samples.

We used a phenomenological spring-dashpot generalized viscoelastic model to characterize the stress relaxation responses of the E and VE hydrogels. This model includes a spring and multiple Maxwell elements that are connected in parallel (Fig. 1F). When utilizing two Maxwell elements, commonly referred to as the Maxwell-Weichert model, we obtain two relaxation times corresponding to each element. The stress in a Maxwell-Weichert model is represented by

$$\sigma = E_\infty \varepsilon_0 + E_1 \varepsilon_0 e^{\left(-\frac{t}{\tau_1}\right)} + E_2 \varepsilon_0 e^{\left(-\frac{t}{\tau_2}\right)} \tag{8}$$

where $\tau_1 = \frac{\eta_1}{E_1}$ and $\tau_2 = \frac{\eta_2}{E_2}$ are the relaxation constants. $E_\infty$, $E_1$, $E_2$ are the moduli for the springs in the model. We fit the stress relaxation data for samples in the E and VE hydrogel groups to the Weichert model and obtained the coefficients in equation (8) that were next used in the FE model(Table 2). In ABAQUS, the Prony series form for relaxation modulus is written as

$$E(t) = E_0 \left(1 - \sum_{i=1}^{N} g_i \left(1 - e^{\frac{-t}{\tau_i}}\right)\right) \tag{9}$$

$E_0$ is the instantaneous modulus, given by $E_0 = E_\infty + \sum_{i=1}^{N} E_i$. The dimensionless relaxation moduli are $g_i = \frac{E_i}{E_0}$ that provide a good equivalence between experimentally observed relaxation behaviours and simulations using the FE model (24).

**Cavitation rheology and bubble growth dynamic in hydrogels**

Cavitation rheology was used to study bubble inflation dynamics in E and VE hydrogels, following established methods (15-17). A 150 µm blunt-tipped needle, connected to a 20 cc syringe, was inserted into the hydrogel samples, and air was injected using a syringe pump at controlled inflation rates of 500, 1000, 2000, 4000, 6000 and 8000 µl/min (Fig. 2A). Bubble pressure during inflation was recorded via a pressure sensor (ASDXAVX030PGAA5, Honeywell Sensotec Inc., Columbus, OH, USA) connected to an Arduino UNO board (16). Temporal pressure variations during inflation were obtained, and the critical pressure, $P_c$, was defined as the peak internal pressure just before bubble collapse. This value was then used to calculate the critical energy release rate (Gc) given by (16-18)

$$\frac{P_c}{\sqrt{E}} = \sqrt{\left(\frac{\pi G_c}{3}\right)} \sqrt{\left(\frac{1}{r}\right)} \tag{10}$$

E is the Young's modulus of the gels, and r is the inner radius of the needle (150 µm) in the cavitation experiments.

Bubble formation ahead of the needle tip was recorded at 240 fps using a phone camera (iPhone XI) mounted above the sample with bottom illumination using an LED light source. We segmented the images during bubble growth in the hydrogel using Fiji (NIH ImageJ) and fit these images to an ellipse to quantify the bubble shape during inflation. The ratio of the major to minor axis of the bubble is defined as the aspect ratio (AR). A spherical bubble has an AR =1, whereas elliptical shapes show significant deviations from this value. We also quantified the effective bubble size, $R_{eff}$ using the computed area, A, and perimeter, P, of the bubble, given as:

$$R_{eff} = \frac{2A}{P} \tag{11}$$

$R_{eff}$ and AR variations with time were used to characterize the bubble growth in elastic and viscoelastic hydrogels.

Experimental results from $R_{eff}$ with time, t, were fit to a form similar to the Richards growth model (41, 42) to obtain changes in the effective bubble radius with time as

$$R_{eff}(t) = R_o + R_1(1 - e^{-\frac{t}{\tau}}) \tag{12}$$

$R_o$ is the initial bubble radius (at time t =0), $R_1$ shows changes in length with time, and $\tau$ is time scale for growth. From equation 12, we calculate the growth velocity as

$$\frac{dR_{eff}(t)}{dt} = \frac{R_1}{\tau} e^{-\frac{t}{\tau}} \tag{13}$$

These expressions were used in simulations to model bubble growth in the E and VE hydrogels in our study.

## Finite Element (FE) simulations

### *Uniaxial compression studies*

Experimental results were used to simulate PAAm hydrogel deformation numerically using Abaqus 2022b (SIMULIA, 2022). The hydrogels were modeled as homogeneous, isotropic and incompressible materials. A neo-Hookean hyperelastic material model was used to capture the nonlinear stress-strain responses (Fig. 1E). Simulations used a structured quad-dominated mesh of 8-node linear brick elements (C3D8RH(S)) with reduced integration implemented using a hybrid formulation with constant pressure. The hybrid formulation is essential for modeling incompressible or nearly incompressible materials, whereas reduced integration with hourglass control minimizes volumetric and shear locking artifacts which leads to artificially stiff behavior and inaccurate stress distributions in incompressible materials. To account for large deformations, geometric nonlinearity was enabled (Nlgeom ='On') using a static general step in the simulations. Results using C3D20RH$^{(S)}$ (20-node quadratic brick, reduced integration, hybrid with linear pressure) and C3D8RH produced similar results. We hence used the C3D8RH that offers a balance between solution accuracy and computational cost.

Mechanical properties of the PAAm hydrogels were defined in the neo-Hookean model using the shear modulus ($\mu = 2C_{10}$). A fixed displacement boundary condition was applied to the top surface, imposing a total compressive strain of 78%, whereas the base was constrained using roller supports. Mesh sensitivity analysis was performed using 3674, 18400, 147200, 220700 and 294200 elements. A mesh size of 220700 was selected based on convergence of peak stress values (Fig.S1A, B; Table S1). This mesh was also used with a time step of 0.01 for the stress relaxation simulations (Table S2). The geometry and mesh employed for the compression simulations were next used to simulate stress relaxation in ABAQUS. A 'Static, General' step was created to apply a compressive strain of 10% to the top surface of the hydrogel over ~ 1 minute and a 'Visco' step over 4 hours was next introduced to analyze stress relaxation of the hydrogel. The nonlinear material properties were obtained using coefficient $C_{10}$ from the uniaxial compression experiments. Viscoelastic parameters for both elastic and viscoelastic PAAm hydrogels were obtained by fitting the simulation outputs to experimental stress-relaxation data using the lsqnonlin function in MATLAB R2023a to determine the coefficients to the Prony series ($g_1(t)$, $g_2(t)$, $\tau_1(s)$, $\tau_2(s)$), calculated using Eq. (9) and listed in Table 2. A reference point was selected on the top surface of the hydrogel for both the compression and stress relaxation simulations. Variations of the reaction force with time were extracted at the reference point, and the engineering stresses were estimated using the reaction forces and the cross sectional areas(Fig. S2, Fig. 1G).

### *Bubble expansion in PAAm hydrogels*

Hyperelastic and viscoelastic material properties, corresponding to the neo-Hookean model and the Prony series coefficients, respectively (Tables 1, 2), were used to simulate 2D bubble expansion in PAAm hydrogels using ABAQUS. A cylindrical geometry, with dimensions similar to the experimentally tested hydrogel sample, was prepared with an initial circular cavity at the centre of the sample. A 2D axisymmetric model, representing half of the specimen, was considered to reduce computational costs. Quadrilateral and hybrid elements (CAX4H, in ABAQUS notation) using spherical coordinates were used in these simulations. The specimen was constrained horizontally along the symmetry axis using roller supports and was free along the lateral circumference. The radial displacement history due to cavity expansion from experimental measurements were used as inputs to the model for all inflation rates in the E and VE hydrogel groups. The time corresponding to the initial cavity radius was taken as the reference initial time (t = 0) and a tabular amplitude (AMP) was used to prescribe the time-dependent radial displacements uniformly on the inner cavity surface in the outward radial direction(Fig. 5A). This approach ensures that the simulated cavity expansion follows the experimentally observed expansion and collapse dynamics.

A reference node was created on the inner surface of the cavity to allow extraction of displacement and stress quantities at the cavity boundary in the hydrogel during the simulation. The number of elements along the circumference was maintained to be constant but was varied along the radial direction. Local seeding along the radial direction was adjusted to obtain an overall aspect ratio < 2, geometric deviation factor < 0.005, with the quad-face corner angle ranging between 90 - 140° to maintain a smooth transition from circular mesh to rectangular shape mesh. We also used a dynamic explicit solver for the step mode. The coordinates were changed to spherical coordinates after convergence and the radial stresses (S11) were plotted at the reference node (Fig. S4A). We also ensured that the computed radial stresses converged for various densities in mesh independence studies (Fig. S4B).

### **Modified Rayleigh-Plesset equation to compute stresses in hydrogels during bubble growth and growth gradients for cavitation-induced bubble dynamics**

We used a mathematical framework, based on a modified form of Rayleigh-Plesset equation, to model the dynamics of bubble growth in the hydrogels (22, 37-39, 43, 44). Growth of a spherical bubble in an infinite, homogeneous, incompressible and viscoelastic medium, whose elasticity is described using a neo-Hookean material, by

$$R\ddot{R} + \frac{3}{2}\dot{R}^2 = \frac{p_B - p_\infty(t)}{\rho} - \frac{4\nu_L \dot{R}}{R} - \frac{2S}{\rho R} - \frac{E_{NH}}{\rho} \tag{14}$$

$R = R(r, t)$ is the bubble radius at a given time, t, $\dot{R}$ and $\ddot{R}$ the time derivatives of the radius. S is the surface tension, $\nu_L$, the viscosity and ρ the hydrogel density. The driving force for bubble evolution is given by the pressure term in equation (14), where $p_B$ is the pressure acting on the inner side of the bubble interface, and $p_\infty(t)$ is the pressure in the medium far away from the bubble taken as the atmospheric pressure.

The viscous stress term is given by $\frac{4\nu_L \dot{R}}{R}$ and the surface tension-based stress is given by $\frac{2S}{R}$ . The elastic stress term, $E_{NH}$, is given in terms of a neo-Hookean material by

$$E_{NH} = \frac{\mu}{2}[5 - 4\left(\frac{R_0}{R}\right) - \left(\frac{R_o}{R}\right)^4] \tag{15}$$

$\mu$ the shear modulus of the hydrogel and $R_0 = R_0(r, 0)$ is the initial bubble radius. The bubble pressure is calculated using (43)

$$p_B = p_v(T_\infty) + p_{Go}\left(\frac{R_o}{R}\right)^{3k} \tag{16}$$

where $p_v$ is the vapor pressure, $T_\infty$, the ambient temperature and $p_{Go}$ is the initial partial pressure of the gas used for inflation, and k is the polytropic index (39). We assumed that the bubble expansion is isothermal, such that k=1.

The dynamics of bubble growth is dependent on the material properties of the medium which determines changes in the bubble radius at the critical pressure, $P_c$, in the cavitation rheology studies. We calculated the shear moduli for the E and VE hydrogels using experimental results from compression tests performed and obtained  coefficient to the neo-Hookean model using the shear modulus as $\mu = 2C_{10}$. The dynamic viscosity ($\eta = \nu_L * \rho$ ) was calculated using Dynamic Mechanical Analyses experiments (23). $\eta = \frac{E''}{\omega}$ where $E''$ is the loss modulus and $\omega$ is the applied frequency. We used values of $G''$ corresponding to a frequency of 8 Hz when the storage modulus $E'$ reaches a steady response. The approximate surface tension (S) values of polyacrylamide for the E and VE hydrogels were taken as 72 mN/mm from literature which is similar to that of water (45). A scaling factor was used to adjust for the reduction in surface tension after gelation as suggested in previous reports (46). A parametric sweep was also performed to show that the effect of surface tension induced stresses were negligible compared to the elastic and viscous stress effects and did not influence the bubble growth dynamics (Fig. S8). Because closed-form analytical solutions to the Rayleigh-Plesset equation are challenging to solve, we used the form of bubble radius growth law to (equation 12) to numerically calculate stresses corresponding to the elastic, viscous and surface tension terms. We also compared these results from this approach for the E and VE hydrogels with those obtained using the FE method.

## Statistical analysis

Results from the two PAAm hydrogel groups were compared using an unpaired t-test. p values were used to determine the level of significance in these comparisons. *** denotes p-value $< 0.001$, ** for $p < 0.01$, and * $p < 0.05$ in this study.

**Acknowledgments**

AS was supported by a PMRF scholarship from the Ministry of Education, government of India. NG is grateful to MoE-STARS (2/2023-0603) and intramural grants from the Institute for project support.

**Author contributions:**

SSP, AS, IJ, SSP - Methodology, Software, Validation, Formal Analysis, Investigation, Data Visualization. AC- Methodology, Software, Validation, Formal Analysis, Investigation, Data Visualization, Writing – Original draft for the modified Rayleigh-Plesset method, Reviewing and Editing. NG: Conceptualization, Methodology, Data Visualization, Formal Analysis, Resources, Data Curation, Writing – Original draft, Writing – Reviewing and Editing, Data Visualization, Supervision, Project Administration and Funding Acquisition.

**Competing interests:**
All authors declare they have no competing interests.

**Data and materials availability:**
All data are available in the main text or the supplementary materials.

## Figures and Tables

**Fig. 1**: **Uniaxial compression and stress relaxation mechanical experiments of the elastic (E) and viscoelastic (VE) hydrogels in the study**. (**A)** Cartoon shows E gels and (**B)** VE gels that included linear PAAm within the material during fabrication. (**C**) A neo-Hookean constitutive model was used to fit the experimentally obtained uniaxial compression results for samples in the E group, and (**D**) VE hydrogels, respectively. Plots show average values (symbols) and the shaded regions show the standard deviation (n=4 in E and VE groups) in these plots. $r^2$ values show goodness of fit between the experimental results and model. (**E**) A finite element (FE) mesh was used to fit the quasi-static compression experiments based on a neo-Hookean model. (**F**) The Weichert model, consisting of a spring in combination with two Maxwell elements, was used to describe the viscoelastic properties of E and VE hydrogels. (**G**) Experimental stress relaxation results are shown for a representative sample in the E group and VE hydrogels using the FE model with Prony series coefficients. The goodness of fits was determined using the $r^2$ values.

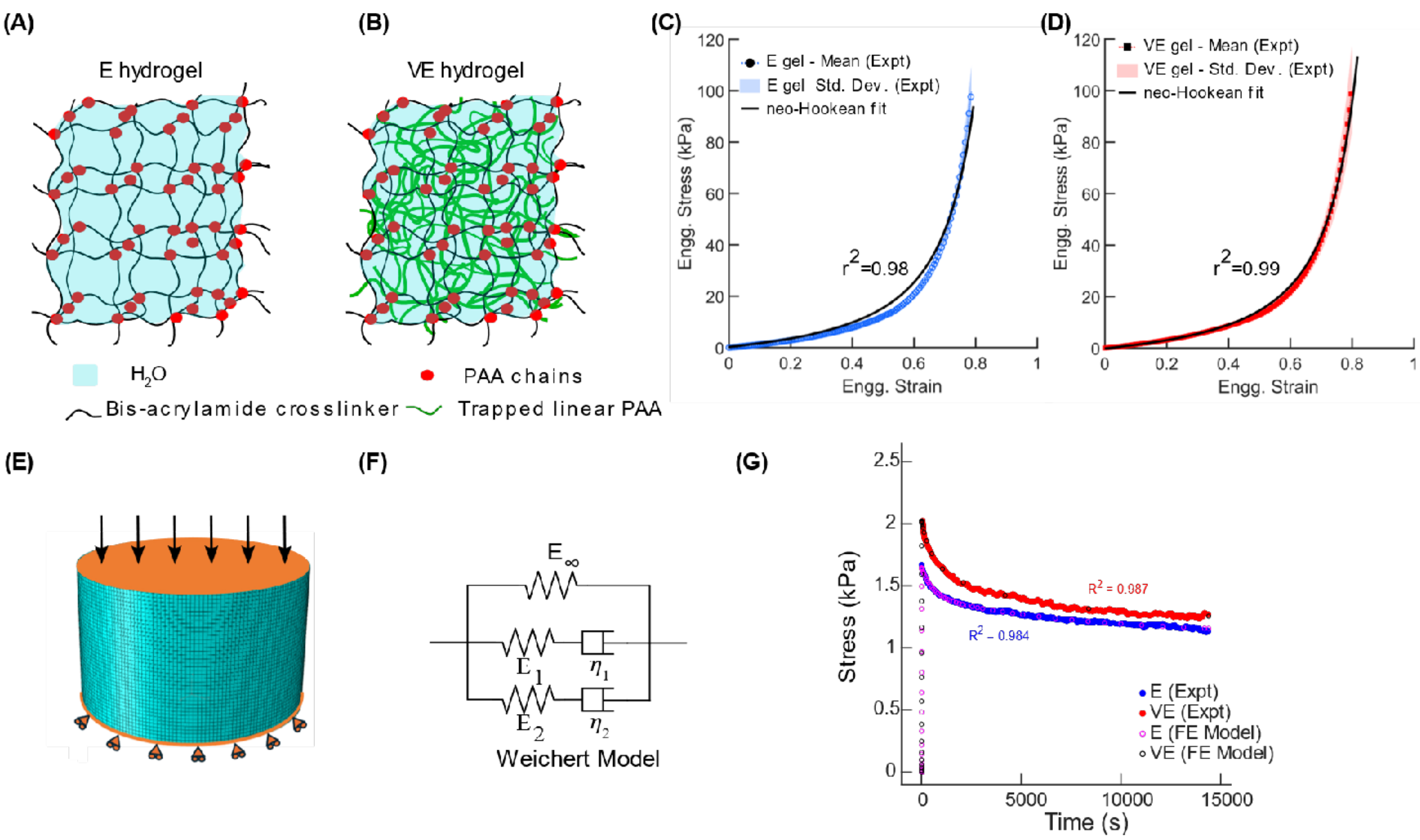

**Fig. 2: Cavitation rheology studies to quantify hydrogel failures. (A)** Cartoon shows the cavitation rheology setup to measure pressure using a transducer and the dynamics of bubble expansion during inflation. **(B)** Representative pressure variations with time are shown for a sample in the E and VE hydrogel groups, respectively, at 2000 µl/ min. The points of cavity initiation and collapse are indicated in the figure. **(C)** Results from cavitation experiments are shown from a representative sample in the E and VE hydrogels for experiments performed at 500, 1000, 2000, 4000, 6000 and 8000 µl/ min to compare differences in bubble growth. **(D)** The bubble initiation time varied inversely with inflation rate for both the E and VE hydrogels. **(E)** The energy release rate, $G_c$, of the E (n=3) and VE (n=3) groups were calculated using critical pressure, $P_c$, and moduli of hydrogels (Equation 10) for each inflation rate. There were no significant differences in $G_c$ for the E and VE groups at all inflation rates. However, $G_c$ was higher for VE gels compared to E gels at each inflation rate in the study ($p<0.05$).

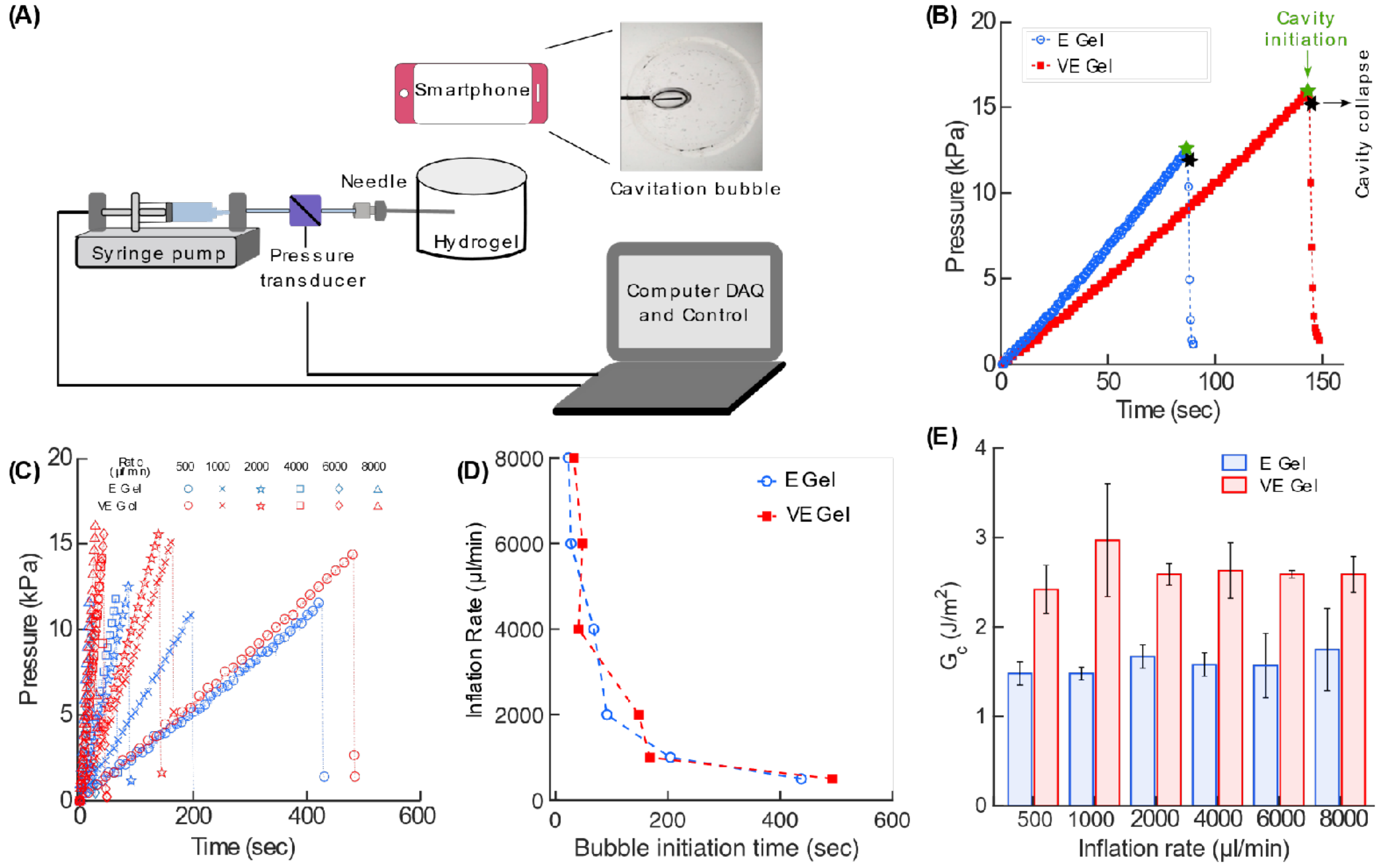

**Fig. 3: Bubble dynamics during cavitation rheology (A)** Cavity shapes are shown just before collapse for representative samples at each of the different inflation rates in the E gel group, and **(B)** VE groups, respectively. **(C)** The bubble edges were tracked and variations in the effective radii ($R_{eff}$) are shown with time for one representative inflation experiment from each of the E and VE hydrogels. **(D)** The corresponding aspect ratios (AR) clearly demonstrate the presence of a minimum for E group compared to monotonic decrease in AR for the VE group. **(E)** The elasto-adhesive, and (F) the elasto-cohesive lengths variations are shown for the E and VE gel groups with inflation rates in the study.

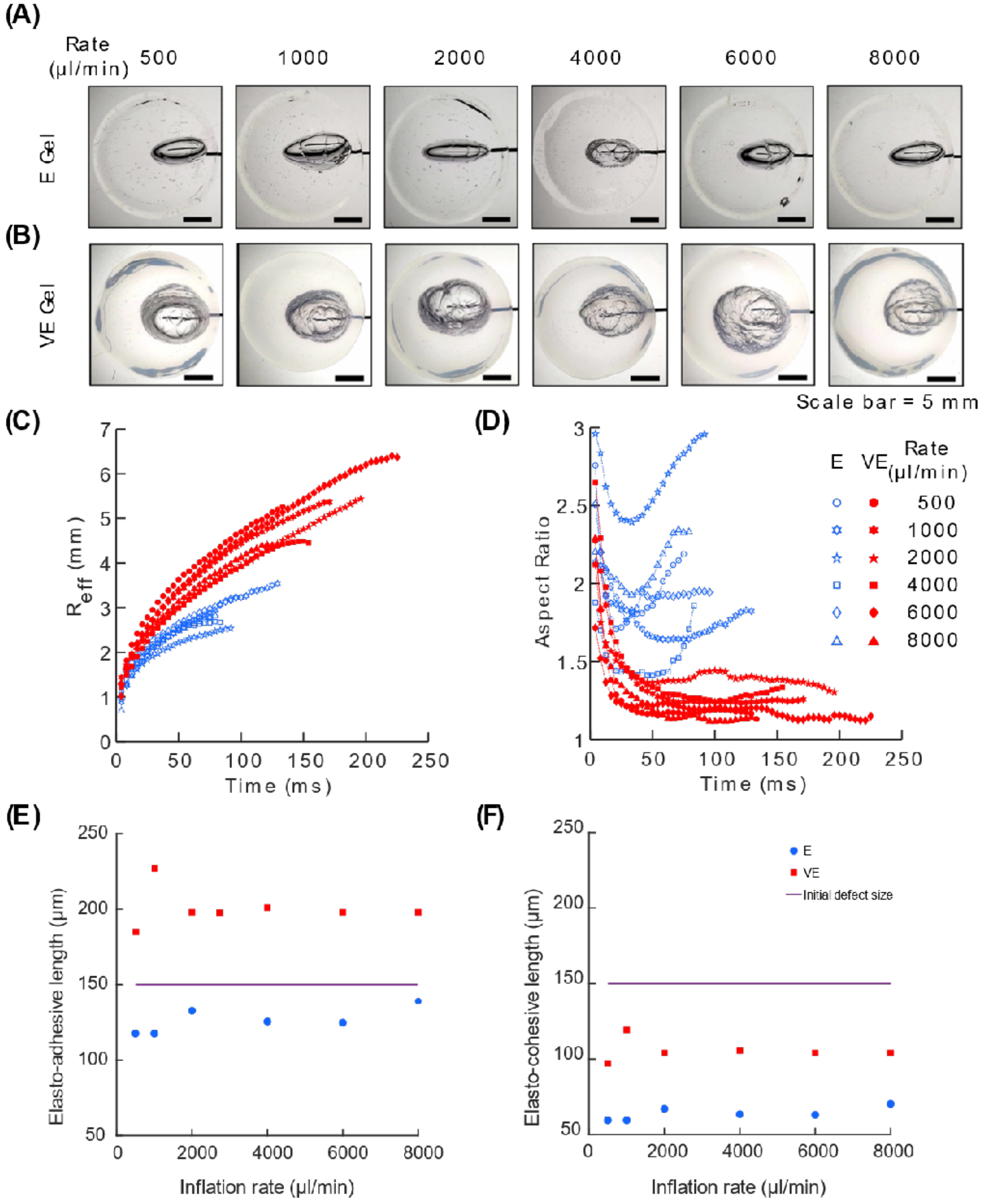

**Fig. 4: Growth of bubbles and the modified Rayleigh-Plesset equation. (A)** Variations in the effective radius ($R_{eff}$) with time are shown for bubble growth in a representative sample from the E and VE hydrogel groups at inflation rates of 500 µl/ min and **(B)** 8000 µl/ min, respectively. The corresponding fits to the Richards growth model (equation 12) are shown on the same plots. **(C)** Radial time dependent growth velocities ($\frac{dR(t)}{dt}$) were calculated for experimental data corresponding to 500 µl/ min and **(D)** 8000 µl/ min inflation rates. These data were fit to the model (equation 13) and are shown on the same plots.

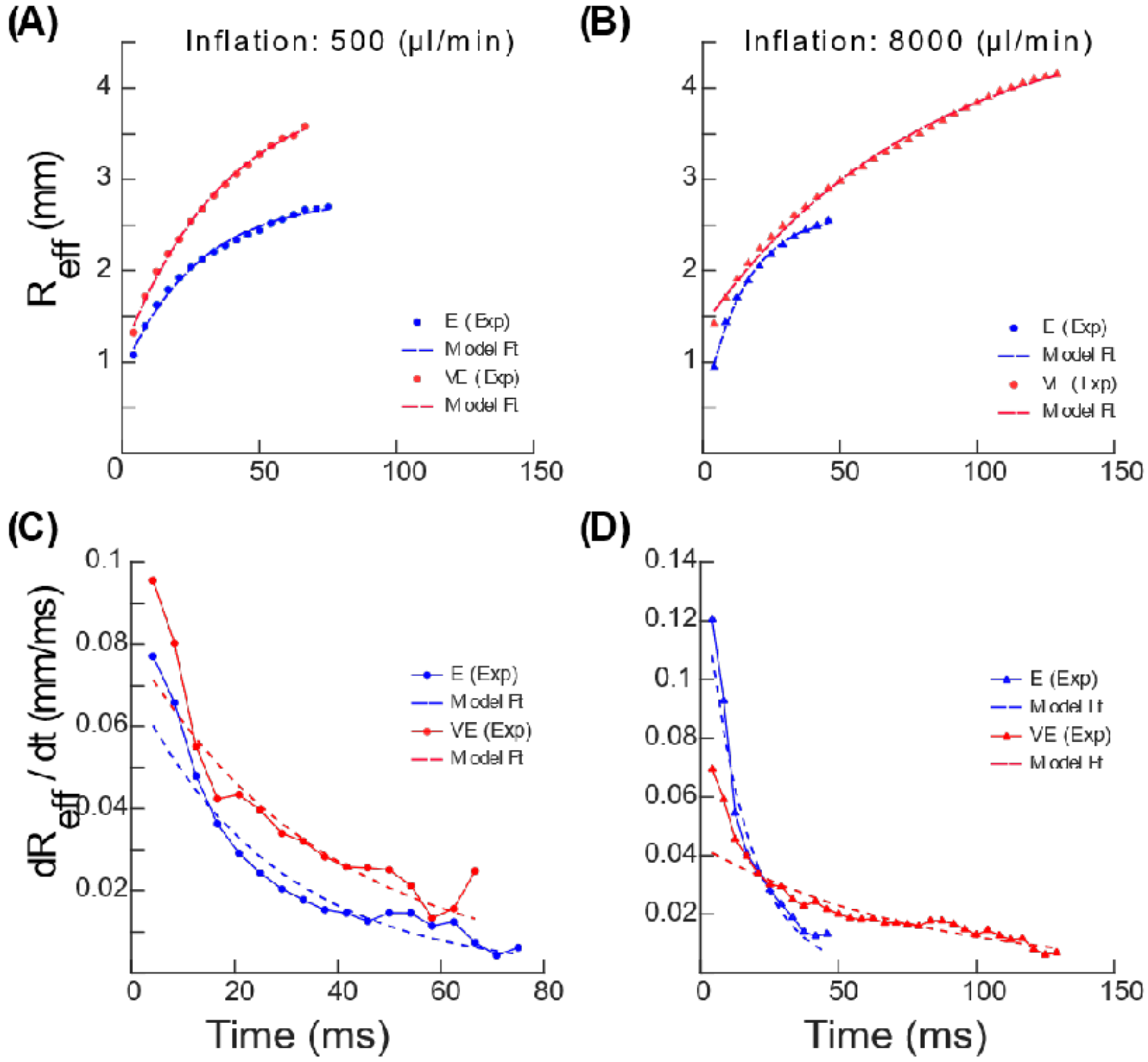

**Fig. 5. The elastic and viscous stresses in E and VE hydrogels during bubble inflation from FE simulations. (A)** A 2D finite element (FE) model was used to simulate bubble growth in the hydrogels. The grades mesh shows smooth transition from circular to rectangular mesh (inset). The cavity growth was simulated using radius changes with time in E and VE gels at various inflation rates. **(B)** The elastic and viscous stresses, calculated using the (analytical) modified Rayleigh-Plesset equation (14), are shown for 500 µl/ min inflation rate in the E hydrogels. Results from FE simulations show a good match with the analytical model. **(C)** The corresponding results are shown for the VE hydrogels with cavity growth at 500 µl/ min. **(D)** Results from the analytical and FE models are shown for the E hydrogels, and **(E)** VE gels at an inflation rate of 8000 µl/ min. **(F)** Comparisons of the resultant growth velocity gradient (GVG), calculated using the modified Rayleigh - Plesset equation, are shown at 500 µl/ min and **(G)** 8000 µl/ min inflation rates. Superposed on these plots are lines corresponding to the inflection point in the elastic stresses (black) obtained from experiments.

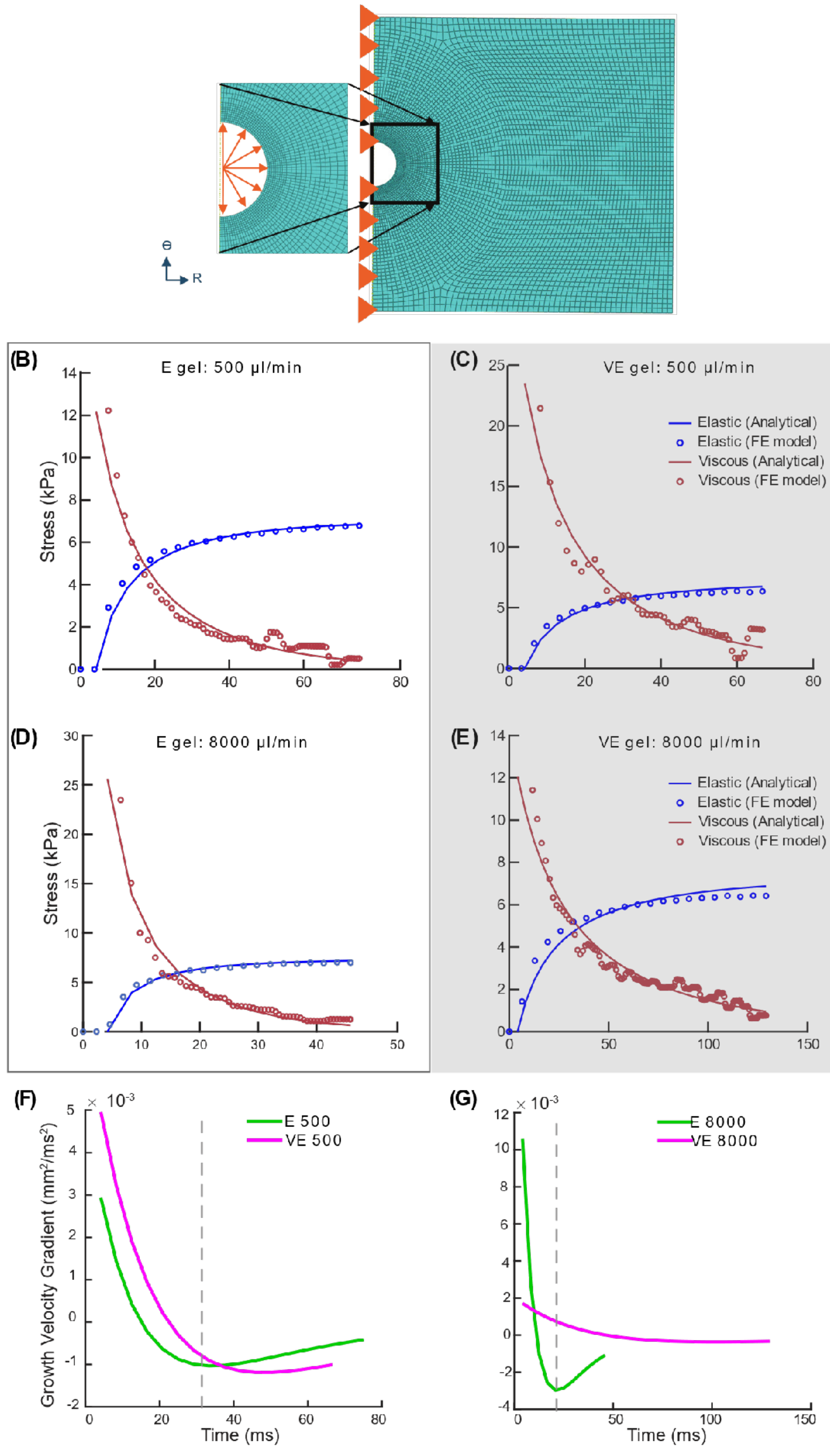

**Fig. 6: Variations in aspect ratio (AR) with Deborah number (De).** Data points #1, #2, and #3 denote the three independent batches tested across all inflation rates. De and AR were calculated at the inflection time for the E gels and at the time when AR reached a constant value for the VE gels.

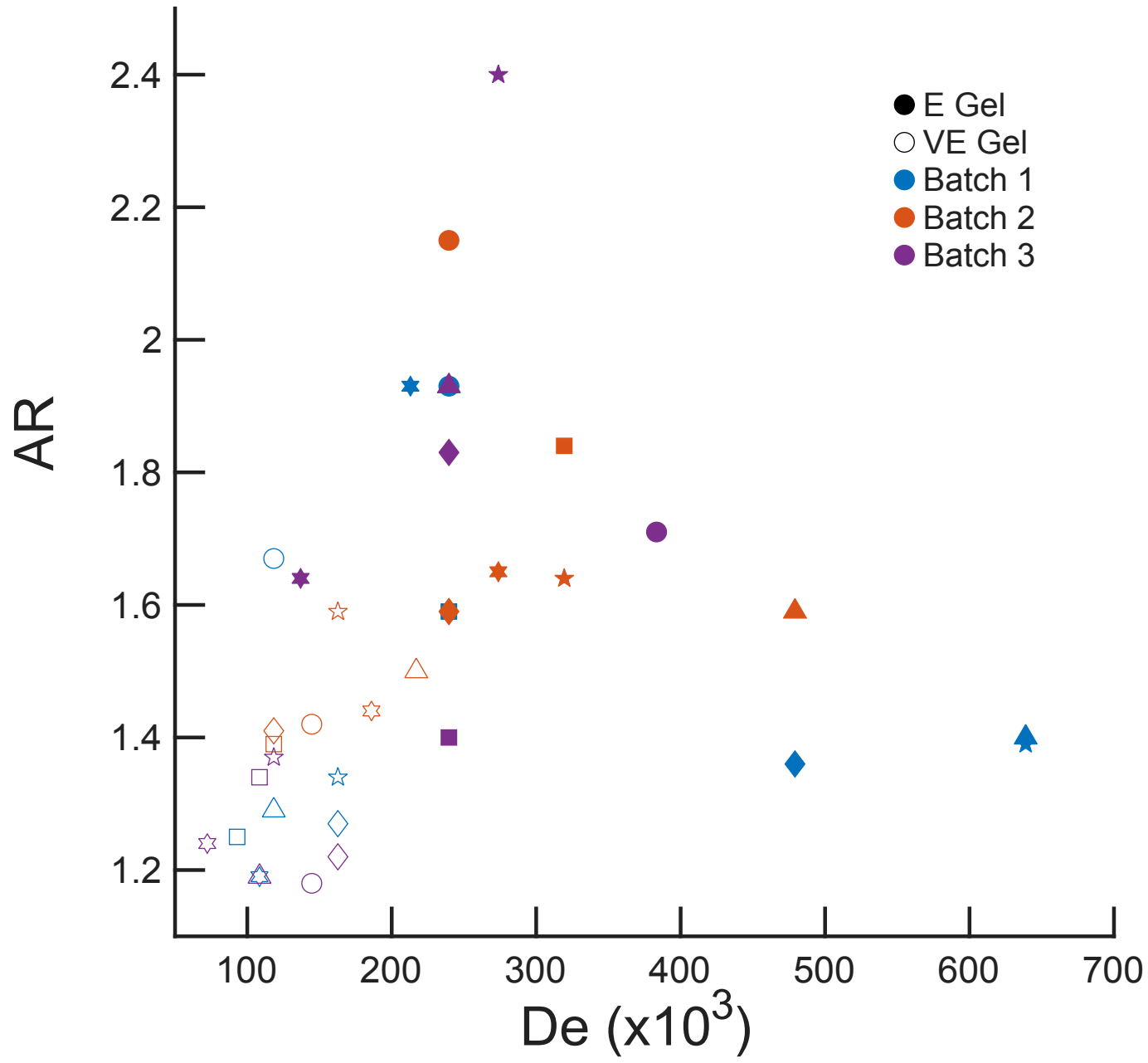

**Table 1: Compression experiment fits to the neo-Hookean model:** Constants to the nonlinear strain energy model were obtained by fitting data from uniaxial compression results for specimens in the E and VE hydrogel group in the study using equation (6).

| **Sample** | **E** | | **VE** | |
|---|---|---|---|---|
| | $C_{10}$ **(kPa)** | $\mathbf{r^2}$ | $C_{10}$ **(kPa)** | $\mathbf{r^2}$ |
| 1 | 2.12 | 0.981 | 2.06 | 0.997 |
| 2 | 2.13 | 0.983 | 1.84 | 0.999 |
| 3 | 2.42 | 0.978 | 2.05 | 0.998 |
| 4 | 1.71 | 0.999 | 1.97 | 0.998 |

**Table 2: Coefficients to the Prony series coefficients.** Stress relaxation results from E and VE gels were fit to the Maxwell-Weichert model and the corresponding coefficients to the two series Prony model were obtained using equation 9.

| **Sample** | $g_1(t) = \frac{E_1}{E_0}$ | $g_2(t) = \frac{E_2}{E_0}$ | $\tau_1$ **(min)** | $\tau_2$ **(min)** | $\mathbf{r^2}$ |
|---|---|---|---|---|---|
| **E** | 0.14 | 0.19 | 9.8 | 133.1 | 0.99 |
| **VE** | 0.19 | 0.21 | 9.4 | 90.4 | 0.99 |

**Table 3: Failure of hydrogels and bubble dynamics.** Results from cavitation rheology experiments on hydrogel samples in the E (n = 3) and VE (n = 3) groups are shown for the different inflation rates used in the study. For the E gels, the mean and standard deviation of the bubble inflection time, determined from the AR–time plots and the GVG_min time (time corresponding to the minimum GVG value), are reported from camera images acquired during inflation.

| **Inflation rate (μl/min)** | **$P_c$ (kPa)** | | **$G_c$ (J/m$^2$)** | | **Bubble inflection time (ms)** | **$GVG_{min}$ Time (ms)** |
|---|---|---|---|---|---|---|
| | **E** | **VE** | **E** | **VE** | **E** | |
| **500** | 11.39 ±0.49 | 14.85 ± 0.85 | 1.48 ± 0.13 | 2.42 ± 0.27 | 29.2 ± 7.2 | 33.33 |
| **1000** | 11.39 ± 0.27 | 16.42 ± 1.57 | 1.48 ± 0.07 | 2.97 ± 0.63 | 43.1 ± 17.4 | 33.33 |
| **2000** | 12.10 ± 0.49 | 15.39 ± 0.36 | 1.67 ± 0.13 | 2.59 ± 0.12 | 23.6 ± 10.5 | 33.33 |
| **4000** | 11.78 ± 0.48 | 15.48 ± 0.89 | 1.58 ± 0.13 | 2.63 ± 0.31 | 44.4 ± 26.8 | 37.50 |
| **6000** | 11.70 ± 1.34 | 15.40 ± 0.13 | 1.57 ± 0.36 | 2.59 ± 0.04 | 27.8 ± 9.6 | 37.50 |
| **8000** | 12.34 ± 1.60 | 15.39 ± 0.60 | 1.75 ± 0.46 | 2.59 ± 0.20 | 20.8 ± 11 | 20.83 |

**Table 4: Constants to the modified Rayleigh - Plesset equation.** $R_{min}$ denotes the minimum bubble radius required for cavitation, calculated using the Wright-Omega functions, which correlated positively with the initial bubble radius $R_o$. Results from model fits to E and VE gels had an $r^2$ >0.8.

| **μl/ min** | **$R_0$ (mm)** | | **$R_1$ (mm)** | | **τ (ms)** | | **$R_{min}$(mm)** | |
|---|---|---|---|---|---|---|---|---|
| | **E** | **VE** | **E** | **VE** | **E** | **VE** | **E** | **VE** |
| **500** | 0.87 | 1.08 | 1.93 | 2.96 | 27.58 | 37.05 | 0.95 | 1.15 |
| **1000** | 0.75 | 1.26 | 2.06 | 3.67 | 23.69 | 67.83 | 0.78 | 1.26 |
| **2000** | 0.81 | 1.14 | 2.04 | 3.48 | 26.66 | 67.24 | 0.82 | 1.16 |
| **4000** | 0.94 | 1.10 | 2.25 | 3.46 | 30.94 | 56.20 | 0.89 | 1.06 |
| **6000** | 0.98 | 1.10 | 2.26 | 3.56 | 32.93 | 54.55 | 0.98 | 1.02 |
| **8000** | 0.46 | 1.38 | 2.16 | 3.45 | 15.20 | 79.93 | 0.69 | 1.17 |

## Supplementary Figures

**Fig. S1 Parametric study for mesh size in FE simulations. (A)** Variations in the maximum compressive stress are shown for the different number of mesh elements for the E hydrogels in the study. **(B)** Images show the meshes assessed in the study. The M4 mesh was used in all simulations reported in this study.

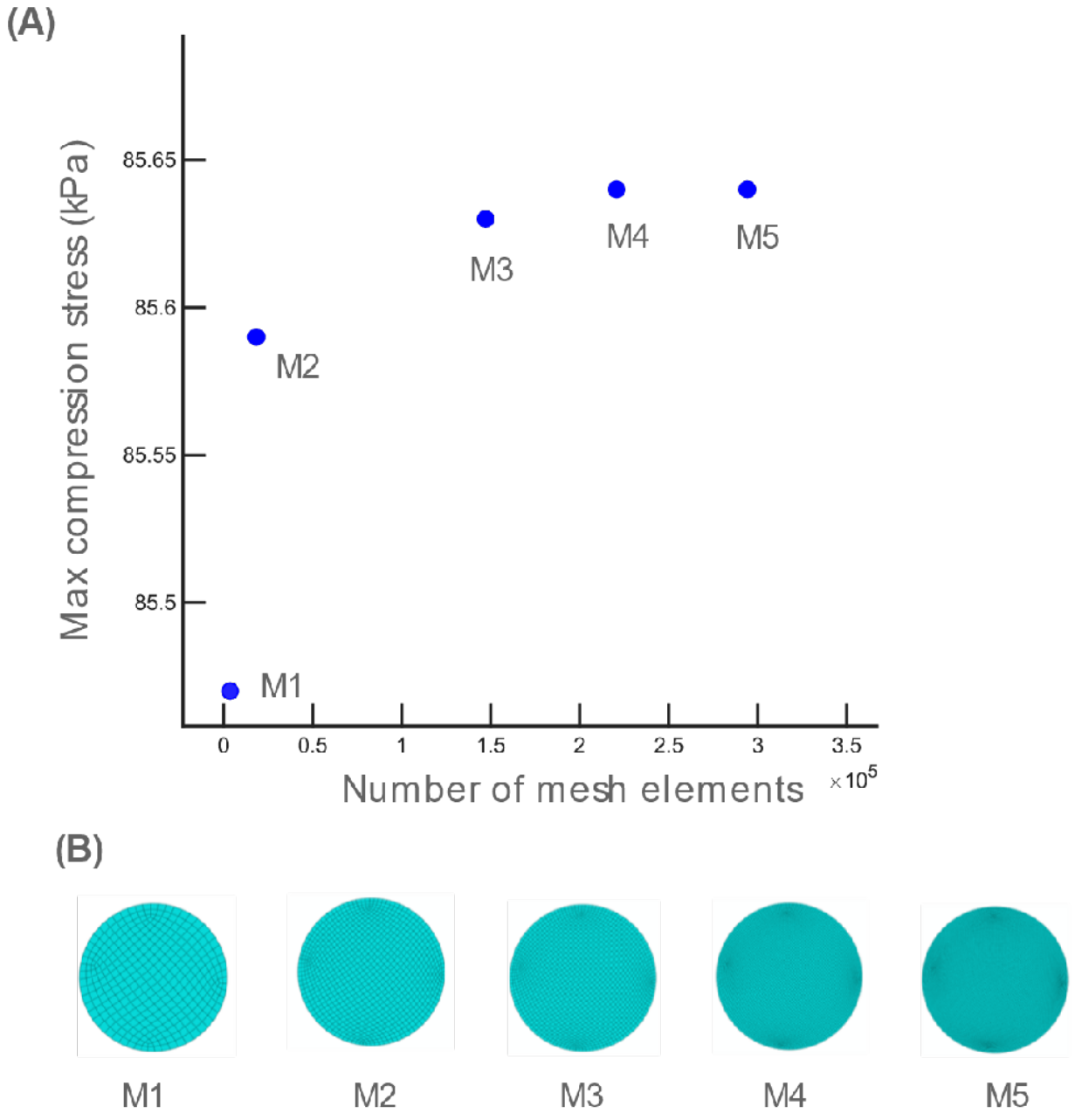

**Fig. S2: Comparisons of FE simulations with monotonic compression experiments. (A)** Plot shows experimental results with simulations using neo-Hookean model coefficient for a representative sample in the E and **(B)** VE hydrogel groups, respectively. $r^2$ values show goodness of fit between the experimental data and the model.

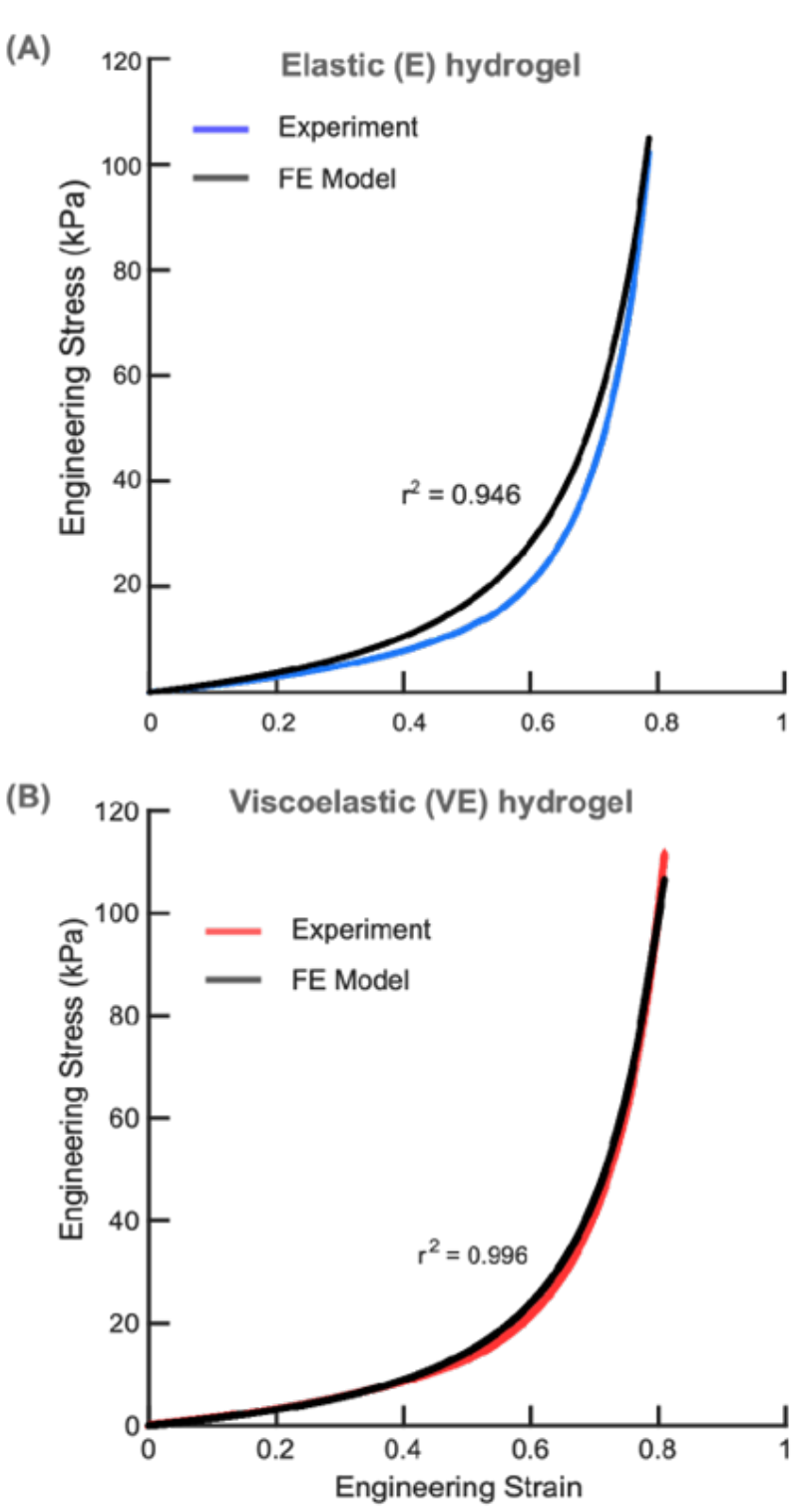

**Fig. S3. Growth of bubbles in E and VE hydrogels.** $R_{eff}$ variations with time are shown for **(A)** E hydrogel and **(B)** VE gels for various inflation rates tested in the study. Experimental data were fit to an exponential Richards growth model (Equation 12) and the goodness of fit was determined using $r^2$ values. The corresponding variations in $dR_{eff}/dt$ are shown for representative samples in the **(C)** E hydrogels, and **(D)** VE hydrogels, respectively, with model fits (Equation 13).

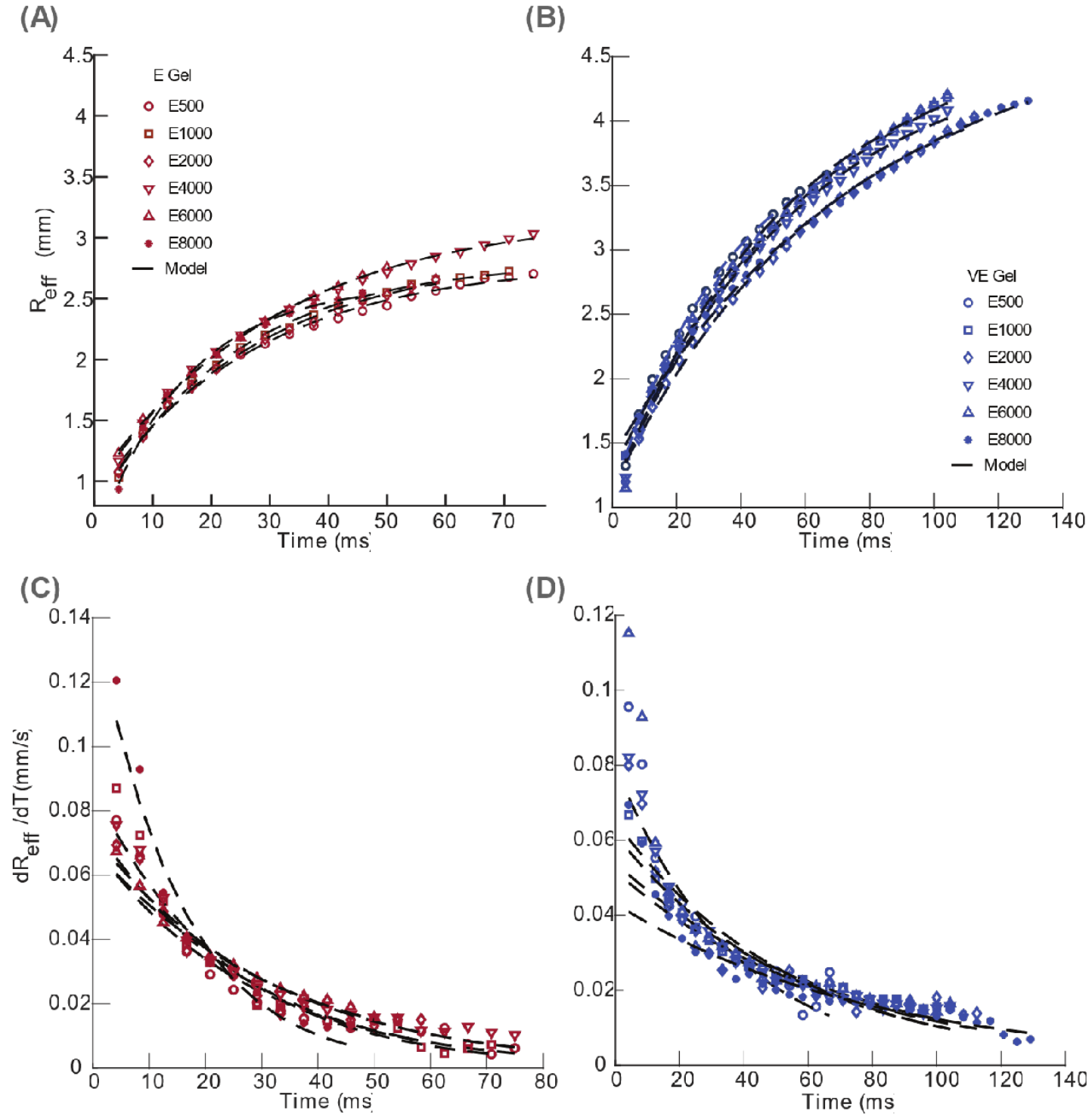

**Fig. S4. Bubble growth dynamics in FE simulations (A)** Radial stress variations in the FE model are shown for the E8000 hydrogel. The mesh near the cavity and regions away from the mesh are also shown. **(B)** A mesh convergence analysis shows variation in radial stresses for different element numbers in the FE model.

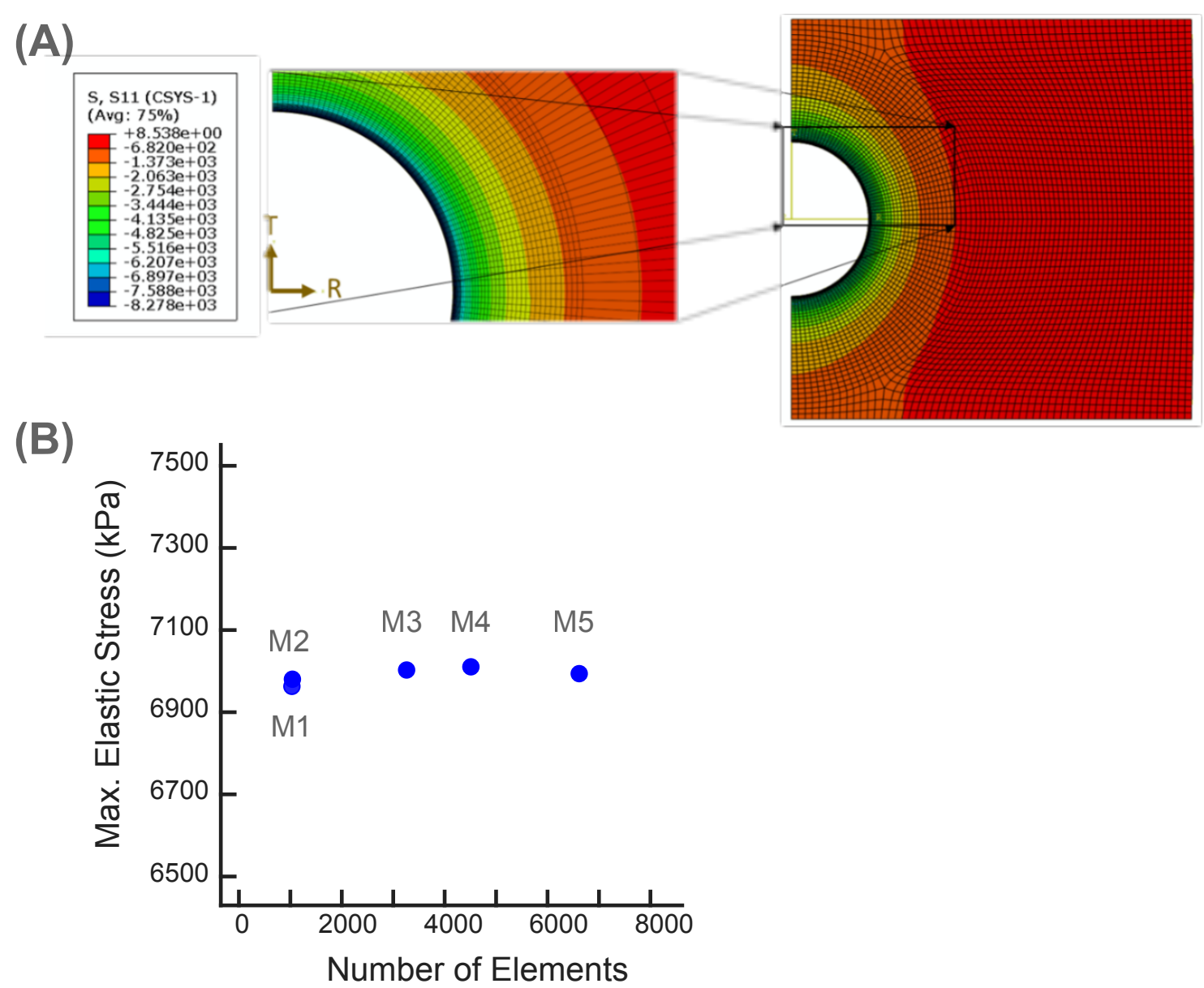

**Fig. S5. Comparison of Aspect Ratios (AR; symbols) and growth velocity gradients (GVG; straight line) with time for the E hydrogels. (A)** Results from experiments and the model law are shown for 500 µl/ min **(B)** 1000 µl/ min, **(C)** 2000 µl/ min, **(D)** 4000 µl/ min, **(E)** 6000 µl/ min, and **(F)** 8000 µl/ min inflation rates in the study.

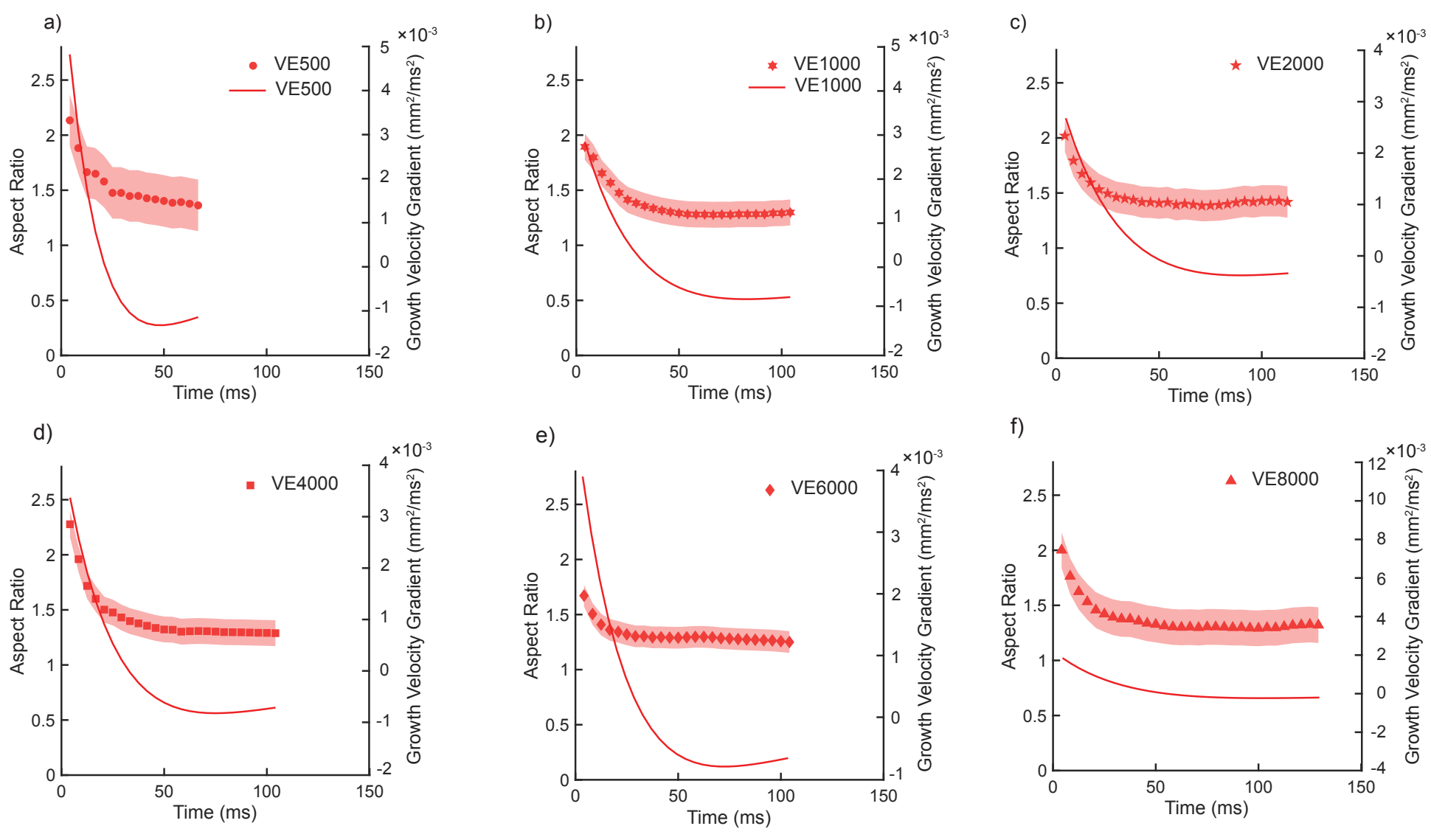

**Fig. S6. Comparison of Aspect Ratios (AR) and growth velocity gradients (GVG) with time for the VE hydrogels. (A)** Results from experiments and the model law are shown for 500 μl/ min **(B)** 1000 μl/ min, **(C)** 2000 μl/ min, **(D)** 4000 μl/ min, **(E)** 6000 μl/ min, and **(F)** 8000 μl/ min inflation rates in the study.

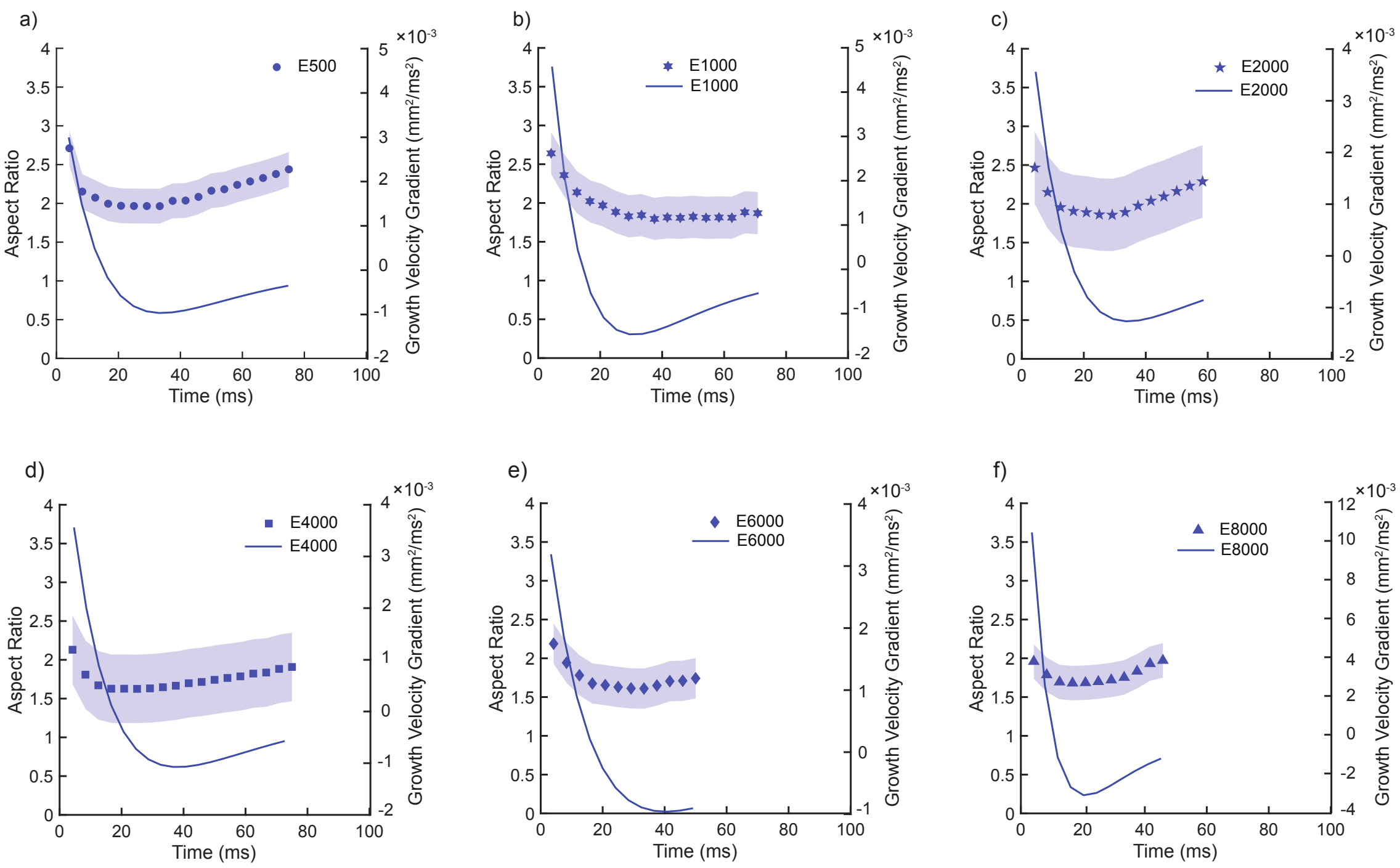

**Fig. S7. Comparison between the modified Rayleigh -Plesset model and FE models for the E and VE hydrogels. (A)** The elastic and viscous stresses are shown for 1000 µl/ min, **(B)** 2000 µl/ min, **(C)** 4000 µl/ min, and **(D)** 6000 µl/ min inflation rates tested in the experiments.

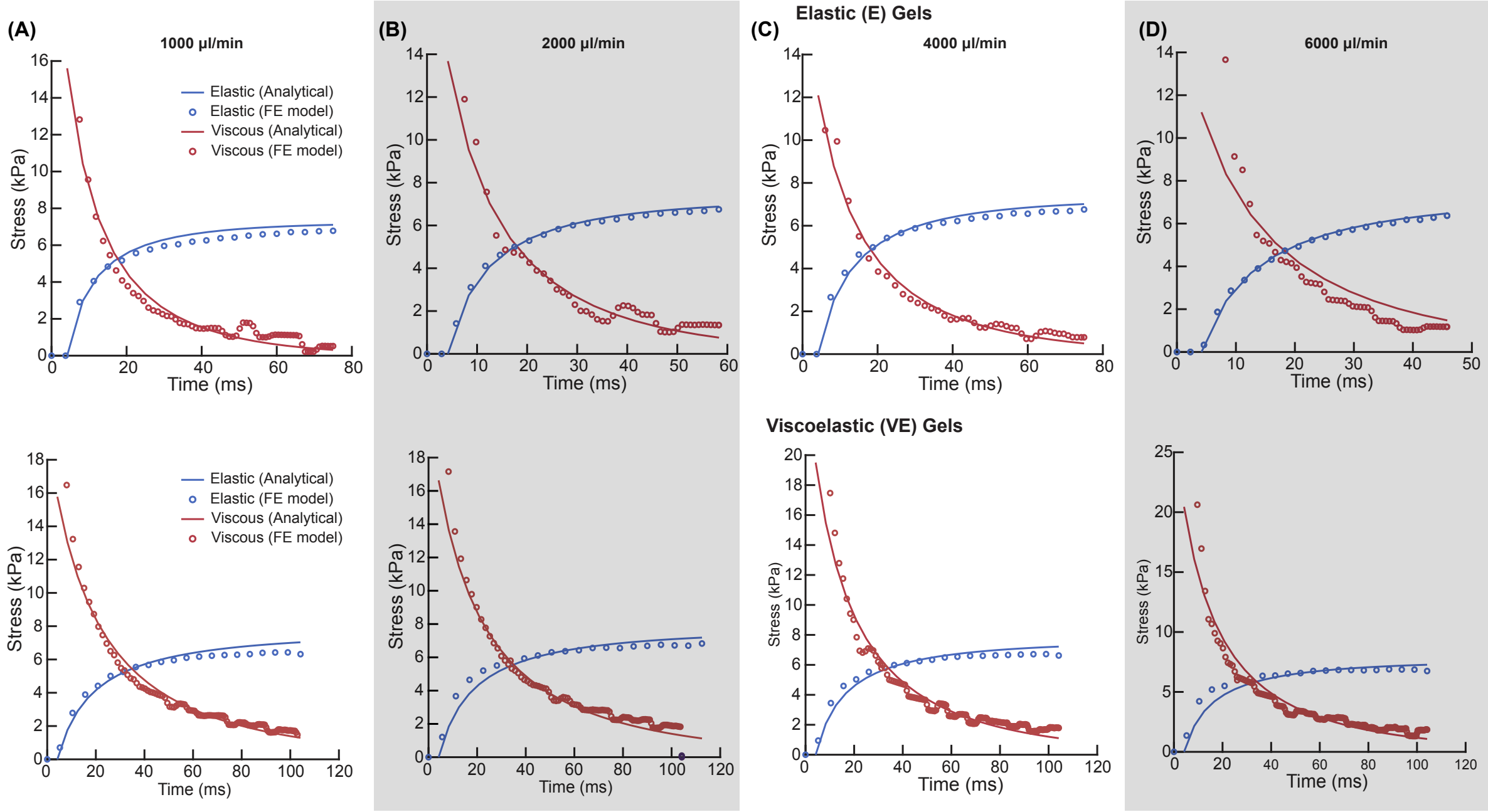

**Fig. S8. Surface tension driven changes in hydrogels.** Surface tension of the hydrogels was varied to test sensitivity of the transition times in the elastic and viscous stresses.

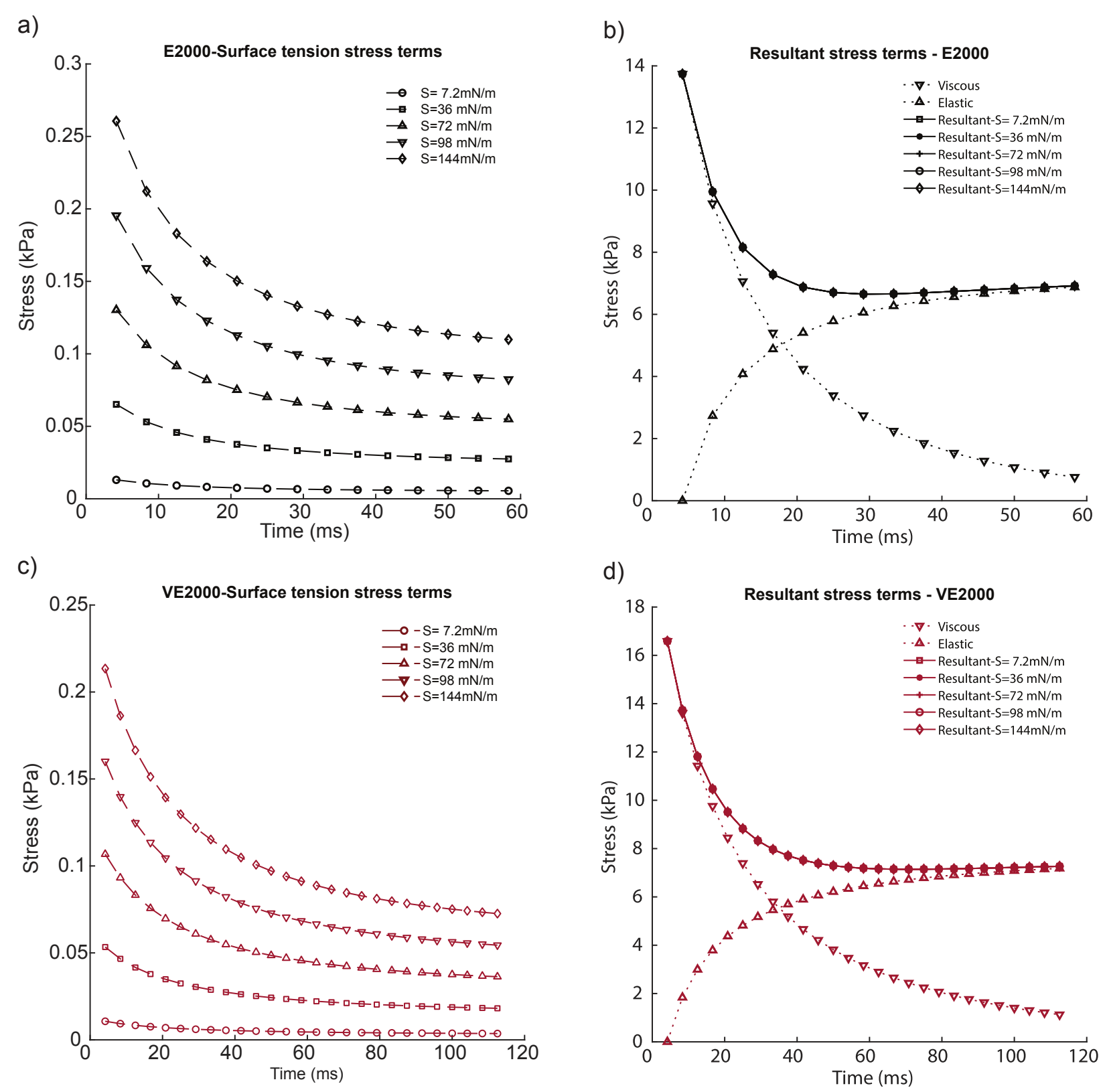

**Supplementary Tables**

**Table S1.** Mesh independence studies for element size used in the finite element simulations.

| **Mesh Type** | **M1** | **M2** | **M3** | **M4** | **M5** |
|---|---|---|---|---|---|
| **# Elements** | 3674 | 18400 | 147200 | 220700 | 294200 |
| **Stress (kPa)** | 85.47 | 85.59 | 85.63 | 85.64 | 85.64 |

**Table S2. Time independence studies using the finite element simulations are shown for the mesh size M4.**

| **SI. No.** | **Initial step time** | **Minimum Increment** | **Maximum Increment** | **Stress (kPa)** |
|---|---|---|---|---|
| **1** | 0.1 | 1E-07 | 1 | 85.63 |
| **2** | 0.01 | 1E-20 | 1 | 85.63 |
| **3** | 0.001 | 1E-20 | 0.005 | 85.63 |

**Supplementary Movies**

**Movie S1: Bubble inflation in elastic hydrogels at 500 μl/ min inflation rate.**

**Movie S2: Bubble inflation in elastic hydrogels at 1000 μl / min inflation rate.**

**Movie S3: Bubble inflation in elastic hydrogels at 2000 μl / min inflation rate.**

**Movie S4: Bubble inflation in elastic hydrogels at 4000 μl / min inflation rate.**

**Movie S5: Bubble inflation in elastic hydrogels at 6000 μl / min inflation rate.**

**Movie S6: Bubble inflation in elastic hydrogels at 8000 μl / min inflation rate.**

**Movie S7: Bubble inflation in viscoelastic hydrogels at 500 μl / min inflation rate.**

**Movie S8: Bubble inflation in viscoelastic hydrogels at 1000 μl / min inflation rate.**

**Movie S9: Bubble inflation in viscoelastic hydrogels at 2000 μl / min inflation rate.**

**Movie S10: Bubble inflation in viscoelastic hydrogels at 4000 μl / min inflation rate.**

**Movie S11: Bubble inflation in viscoelastic hydrogels at 6000 μl / min inflation rate.**

**Movie S12: Bubble inflation in viscoelastic hydrogels at 8000 μl / min inflation rate.**